\documentclass[a4paper]{article}

\usepackage{amsfonts,amssymb,latexsym}
\def\v{\vec}
\def\pa{\partial}
\def\va{\varphi}
\def\la{\lambda}
\def\iy{\infty}
\def\De{\Delta}
\def\de{\delta}

\def\vk{\varkappa}
\def\tvk{\tilde\varkappa}
\def\lan{\langle}
\def\ran{\rangle}
\def\CP{{\mathcal P}}
\def\CH{{\mathcal H}}

\def\CC{{\mathcal C}}

\def\BR{{\mathbb R}}
\def\BZ{{\mathbb Z}}
\def\BC{{\mathbb C}}
\def\BS{{\mathbb S}}
\def\FH{{\mathfrak H}}
\def\Sp{\mathop{\rm Sp}\nolimits\,}
\def\Im{\mathop{\rm Im}\nolimits\,}
\def\Re{\mathop{\rm Re}\nolimits\,}
\def\const{\mathop{\rm const}\nolimits\,}
\def\diag{\mathop{\rm diag}\nolimits\,}
\def\Arg{\mathop{\rm Arg}\nolimits\,}
\def\h{\hbar}
\def\al{\alpha}
\def\ch{\mathop{\rm ch}\nolimits\,}
\def\sh{\mathop{\rm sh}\nolimits\,}
\def\dac{\displaystyle\frac}
\def\dum{\displaystyle\sum}
\def\inl{\int\limits}
\def\dil{\displaystyle\int\limits}
\def\{{\lbrace}
\def\}{\rbrace}
\newtheorem{assum}{Supposition}
\newtheorem{lem}{Lemma}
\newtheorem{demo}{Statement}
\newtheorem{rem}{Remark}
\newtheorem{teo}{Theorem}

\newtheorem{defin}{Definition}

\newtheorem{sled}{Corollary}

\makeatletter
\@addtoreset{equation}{section}
\@addtoreset{teo}{section}
\@addtoreset{lem}{section}
\@addtoreset{demo}{section}
\@addtoreset{rem}{section}
\@addtoreset{prop}{section}
\@addtoreset{sled}{theo}
\makeatother

\title{The Trajectory-Coherent Approximation and the System
of Moments for the Hartree Type Equation}

\bigskip
\author{V.V. Belov$^1$\thanks{e-mail: belov@amath.msk.ru},
A.Yu. Trifonov$^2$\thanks{e-mail: trifonov@phtd.tpu.edu.ru},
and A.V. Shapovalov$^3${\thanks{e-mail:shpv@phys.tsu.ru}}}

\date{
$^1$ Moscow Institute of Electronics and Mathematics\\
3/12 Trekhsvyatitel'sky Lane, 109028 Moscow\\
$^2$ Tomsk Polytechnic University\\
30 Lenin Ave., 634034 Tomsk\\
$^3$ Tomsk State University\\
36 Lenin Ave., 634050 Tomsk}

\begin{document}

\maketitle

\begin{abstract}
The general construction of quasi-classically concentrated solutions to
the Hartree type equation, based on the complex WKB--Maslov method, is
presented. The formal solutions of the Cauchy problem for this
equation, asymptotic in small parameter $\h$ ($\h\to0$), are constructed
with a power accuracy of $O(\h^{N/2})$, where $N$ is any natural number.
In constructing the quasi-classically concentrated solutions, a set of
Hamilton--Ehrenfest equations (equations for centered
moments) is essentially used. The nonlinear superposition principle has
been formulated for the class of quasi-classically concentrated
solutions of Hartree type equations. The results obtained are
exemplified by a one-dimensional Hartree type equation with a Gaussian
potential.
\end{abstract}

\section*{Introduction}

The nonlinear Schr\"odinger equation
\begin{equation}
\{ -i\pa_t +\hat\CH(t,|\Psi|^2)\}\Psi =0, \label{(0.1)}
\end{equation}
where $\hat\CH(t,|\Psi|^2)$ is a nonlinear operator,
arises in describing a broad spectrum of physical phenomena.  In
statistical physics and quantum field theory, the generalized model of
the evolution of bosons is described in terms of the second quantization
formalism by the Schr\"odinger equation \cite{Huang} which, in
Hartree's approximation, leads to the classical multidimensional
Schr\"odinger equation with a nonlocal nonlinearity for one-particle
functions, i.e., a Hartree type equation.

The quantum effects associated with the propagation of an optical pulse in a
nonlinear medium are also described in the second quantization formalism by
the one-dimensional Schr\"odinger equation with a delta-shaped interaction
potential. In this case, the Hartree approximation results in the classical
nonlinear Schr\"odinger equation \cite{Lai1, Lai2} which is integrated by the
Inverse Scattering Transform (IST) method and has soliton solutions
\cite{Zakh1}.  Solitons are localized wave packets propagating without
distortion and interacting elastically in mutual collisions.  The
soliton theory has found wide application in various fields of
nonlinear physics \cite{Zakh, Ablov, Dodd, Newell}.

Investigations of the statistical properties of optical fields have
led to the concept of compressed states of a field in which quantum
fluctuations are minimized and the highest possible accuracy of optical
measurements is achieved.  The important problem of the correspondence
between the stressed states describing the quantum properties of a
radiation and the optical solitons is analyzed in \cite{Lai1, Lai2}.

The Hartree type equation is nonintegrable by the IST
method.  Nevertheless, approximate solutions showing some properties
characteristic of solitons can be constructed.  Solutions of
this type are referred to as solitary waves or ``quasi-solitons'' to
differentiate them from the solitons (in the strict sense) arising in
IST integrable models.

An efficient method for constructing solutions of this type is
offered by the technique of quasi-classical asymptotics.  Thus, for
nonlinear operators of the self-consistent field type, the theory of
canonical operators with a real phase has been constructed for the
Cauchy problem in \cite{Mas1, Mas2} and for spectral problems, including
those with singular potentials, in \cite{Mas3, Karas} (see also
\cite{Bab, Molot, Vacul}). Solitonlike solutions of the Hartree type
equation and some types of interaction potentials have also been
constructed in \cite{Chetv}.

In this paper, localized solutions asymptotical in small
parameter $\h$ ($\h\to0$) for the (nonlinear) Hartree type equation are
constructed using the so-called WKB method or the Maslov complex germ
theory \cite{Mas4, Mas4e, Bagre}.  The constructed solutions are a
generalization of the well-known quantum mechanical coherent and
compressed states for linear equations \cite{Chern, Manko} for the case of
nonlinear Hartree type equations with variable coefficients. We refer to
the corresponding asymptotical solutions, like in the linear case
\cite{Bagre}, as {\em quasi-classically concentrated} solutions
(or states).

Most typical of solitary waves (``quasi-solitons'') is that they
show some properties characteristic of particles.  For the ``quasi-solitons''
being quasi-classically concentrated states of a Hartree type
equation these properties are represented by a dynamic set of
ordinary differential equations for the ``quantum'' means $\v X(t,\h)$
and $\v P(t,\h)$ of the operators of coordinates $\hat x$ and
momenta $\hat p$ and for the centered higher-order moments.  In the
limit of $\h\to 0$, the centroid of such a quasi-soliton
moves in the phase space along the trajectory of this dynamic system:
at each point in time, the quasi-classically concentrated state is
efficiently concentrated in the neighborhood of the point $\v X(t,0)$
(in the $x$ representation) and in the neighborhood of the point
$\v P(t,0)$ (in the $p$ representation).  Note that a similar set of
equations in quantum means has been obtained in \cite{JMPS, BBK} for
the linear case (Schr\"odinger equation) and in \cite{Bagre}
for a more general case. It has been shown \cite{BK1, BK2} that these
equations are Poisson equations with respect to the (degenerate) nonlinear
Dirac bracket.  Therefore, we call the equations in quantum means for
the Hartree type equation, like in the linear case \cite{Bagre},
Hamilton--Ehrenfest equations.  The Hamiltonian character of these
equations is the subject of a special study.  Nevertheless, it should be
noted that, as distinct from the linear case, the construction of the
quasi-classically concentrated states for the Hartree type equation
essentially uses the solutions of the Hamilton--Ehrenfest equations.

The specificity of the Hartree type equation, where nonlinear
terms are only under the integral sign, is that it shows some properties
inherent in linear equations.  In particular, it has been demonstrated
that for the class of quasi-classically concentrated solutions of this
type of equation (with a given accuracy $\h$, $\h\to0$), the {\em
nonlinear superposition principle} is valid.

In terms of the approach under consideration, the formal asymptotical
solutions of the Cauchy problem for this equation and the evolution
operator have been constructed in the class of trajectory-concentrated
functions, allowing any accuracy in small parameter $\h$, $\h\to0$.

It should be stressed that throughout the paper we deal with the
construction of the formal asymptotical solutions to the Hartree type
equation with the residual whose norm has a small estimate in parameter
$\h$, $\h\to0$. To substantiate these asymptotics for finite times
$t\in[0,T]$, $T=\const$, is a special nontrivial mathematical problem.
This problem is concerned with obtaining {\em a priori} estimates for
the solution of nonlinear equation (\ref{(0.1)}), which are uniform
in parameter $\h\in]0,1]$, and is beyond the scope of the present work.
Note that, in view of the heuristic considerations given in \cite{Mas3},
it seems that the difference between the exact and the constructed
formal asymptotical solution can be found with the use of the
method developed in \cite{Mas3, Mas3a}.

This paper is arranged as follows:  The first section gives
principal notions and definitions.  In the second section, a class of
trajectory-concentrated functions is specified and the simplest
properties of these functions are considered.  In the third section,
Hamilton--Ehrenfest equations are constructed which describe the
``particle-like'' properties of the quasi-classically concentrated
solutions of the Hartree type equation.  In the fourth section, the
Hartree type equation is linearized for the solutions of the
Hamilton--Ehrenfest equations, and a set of associated linear equations
which determine the asymptotical solution of the starting problem is
obtained. In the fifth section, we construct, accurate to
$O(\h^{3/2})$, quasi-classical coherent solutions to the Hartree type
equation.  In the sixth section, the principal term of the
quasi-classical asymptotic of this equation is obtained in a class of
quasi-classically concentrated functions. The quasi-classically
concentrated solutions to the Hartree type equation are constructed
with an arbitrary accuracy in $\sqrt\h$ in Sec. 7, while the kernel of
the evolution operator (Green function) of the Hartree type equation is
constructed in Sec. 8.  Herein, the nonlinear superposition principle
is substantiated for the class of quasi-classically concentrated
solutions.  In the ninth section, the Hartree type equation with a
Gaussian potential is considered as an example. Appendix A presents the
properties of the solutions to a set of equations in variations
necessary to construct the asymptotical solutions and the approximate
evolution operator to the Hartree type equation.

\section{The Hartree type equation}

In this paper, by the Hartree type equation is meant the equation
\begin{equation}
\{ -i\h\pa_t +\hat \CH(t)+\vk\hat V(t,\Psi)\}\Psi =0,
\qquad \Psi\in L_2(\BR^n_x). \label{bbst1.1}
\end{equation}
Here, the operators
\begin{eqnarray}
& \hat {\CH} (t)=\CH (\hat z, t), \label{bbst1.2}\\
& \hat V(t,\Psi )= \dil_{{\Bbb R}^n} d\v y\,\Psi^*(\v y ,t)
V(\hat z,\hat w,t)\Psi (\v y ,t) \label{bbst1.3}
\end{eqnarray}
are functions of the noncommuting operators
\[ \hat z=( -i\h\dac\pa{\pa\v x}, \v x), \qquad
\hat w=( -i\h\dac\pa{\pa\v y}, \v y), \qquad
\v x,\v y\in {\Bbb R}^n, \]
the function $\Psi^*$ is complex conjugate to $\Psi$, $\vk$ is a real
parameter, and $\h$ is a ``small parameter'', $\h\in[0,1[$.  For the
operators $\hat z$ and $\hat w$, the following commutative relations are
valid:
\begin{eqnarray}
&\left[ \hat z_k,\hat z_j \right]_- =\left[ \hat
w_k,\hat w_j \right]_- = {i}\h J_{kj},\label{1.kom}\\
& \left[ \hat z_k,\hat w_j \right]_- = 0,\qquad
k,j=\overline {1, 2n},\label{2.kom}
\end{eqnarray}
where $J =\|J_{kj}\|_{2n\times 2n}$ is a unit symplectic matrix
$$ J=\left(\begin{array}{cc}0&-{\Bbb I}\\{\Bbb I}& 0 \end{array}
\right)_{2n\times 2n}.
$$

For the functions of noncommuting variables, we use the Weyl ordering
\cite{Karas2, Groot}.  In this case, we can write, for instance, for the
operator $\hat\CH$:
\begin{eqnarray}
\lefteqn{\hat{\CH}(t)\Psi(\vec x,t,\h)=}\cr
& =\dac1{(2\pi \h)^n}\int\limits_{\BR^{2n}}d\vec yd\vec p
\exp\Big(\frac i\h\lan (\vec x-\vec y),\vec p\ran\Big)
\CH\Big(\vec p,\frac{\vec x+\vec y}{2},t\Big)
\Psi(\vec y,t,\h),\label{bbst1.4}
\end{eqnarray}
where $\CH(z,t)=\CH(\vec p,\vec x,t)$ is the Weyl symbol of the operator
$\hat\CH(t)$ and $\lan .,.\ran$ is the Euclidean scalar product of the
vectors
\[ \lan\vec p,\vec x\ran=\sum^n_{j=1}p_jx_j, \quad
\vec p,\vec x\in \BR^n, \qquad \lan z,w\ran=\sum^{2n}_{j=1}z_jw_j,
\quad z,w\in \BR^{2n}. \]

We here are interested in localized solutions of equation (\ref{bbst1.1})
for each fixed $\h\in[0, 1[$ and $t\in\BR$, belonging to the Schwartz space
with respect to the variable $\v x\in\BR^n$.  For the operators $\hat\CH(t)$
and $\hat V(t,\Psi)$ to be at work in this space, it is sufficient that
their Weyl symbols $\CH(z,t)$ and $V(z, w,t)$ be smooth
functions\footnote{In what follows we assume that for all the operators
under consideration, $\hat A=A(\hat z,t)$, their Weyl symbols satisfy
Supposition \ref{assum1}.} and grow, together with their derivatives,
with $|z|\to\iy$ and $|w|\to\iy$ no more rapidly as the polynomial and
uniformly in $t\in \BR^1$.  Therefore, we believe that the following
conditions for the functions $\CH(z)$ and $V(z, w,t)$ are satisfied:

\begin{assum} \label{assum1}
For any multi-indices $\alpha$, $\beta$, $\mu$, and $\nu$ there exist
constant $C^\alpha_\beta (T)$ and $C^{\alpha\mu}_{\beta\nu}(T)$, such
that the inequalities
\begin{eqnarray*}
& \Big|z^\alpha\dac{\pa^{|\beta|}\CH(z, t)}{\pa z^\beta}\Big| \le
C^\alpha_\beta (T),\\
& \Big| z^\alpha w^\mu\dac{\pa^{|\beta + \nu|}V(z, w,t)}
{\pa z^\beta \pa w^\nu}\Big|\le C^{\alpha\mu}_{\beta\nu}(T) ,\\
& z,w\in {\Bbb R}^{2n} \quad 0\le t\le T
\end{eqnarray*}
are fulfilled.
\end{assum}
Here, $\alpha,\beta,\mu,\nu$ are multi-indices ($\alpha,\beta,\mu,\nu
\in{\Bbb Z}^{2n}_+$) defined as
\begin{eqnarray*}
&\alpha=(\alpha_1,\alpha_2,\dots,\alpha_{2n}), \quad
|\alpha|=\alpha_1+\alpha_2+\dots+\alpha_{2n}, \\
& z^\alpha=z_1^{\alpha_1} z_2^{\alpha_2}\dots  z_{2n}^{\alpha_{2n}},\\
& \dac {\pa^{|\alpha |}V(z)}{\pa z^\alpha}=
\frac{\pa^{|\alpha|}V(z)}{\pa z_1^{\alpha_1}\pa z_2^{\alpha_2}\dots
\pa z_{2n}^{\alpha_{2n}}}, \quad \alpha_j=\overline{0,\iy}, \quad
j=\overline{1,2n}.
\end{eqnarray*}

We are coming now to the description of the class of functions
for which we shall find asymptotical solutions to equation (\ref{bbst1.1}).

\section{The class of trajectory-concentrated functions}

Let us introduce a class of functions singularly depending on a small
parameter $\h$, which is a generalization of the notion of a solitary
wave.  It appears that asymptotical solutions to equation (\ref{bbst1.1})
can be constructed based on functions of this class, which depend on
the phase trajectory $z=Z(t,\h)$, the real function $S(t,\h)$
(analogous to the classical action at $\vk=0$ in the linear case), and
the parameter $\h$.  For $\h\to0$ the functions of this class are
concentrated in the neighborhood of a point moving along a given phase
curve $z=Z(t,0)$.  Functions of this type are well known in quantum
mechanics.  In particular, among these are coherent and ``compressed''
states of quantum systems with a quadric Hamiltonian
\cite{Chern,Manko,Schr,Gla1,Gla2,Rash,Kla1,Kla2,Kla3,Perel}.  Note that the
soliton solution localized only with respect to spatial (but not
momentum) variables does not belong to this class.

Let us denote this class of functions as $\CP_\h^t(Z(t,\h),S(t,\h))$ and
define it as
\begin{eqnarray}
\lefteqn{\CP_\h^t=\CP_\h^t\big(Z(t,\h),S(t,\h)\big) =}\cr
&\displaystyle=\biggl\{\Phi :\Phi (\v x,t,\h)=
\va\Bigl(\frac{\De\v x}{\sqrt{\h}},t,\h\Bigr)
\exp\Bigl[{\frac{i}{\h}(S(t,\h)+
\lan\v P(t,\h),\De\v x \ran )}\Bigr]\biggr\}, \label{bbst1.5}
\end{eqnarray}
where the function $\va(\v\xi ,t,\h)$ belongs to the Schwartz space $\Bbb S$
in variable $\v\xi\in\BR^n$ and depends smoothly on $t$ and
regularly on $\sqrt\h$ for $\h\to0$.  Here, $\De\v x=\v x-\v X(t,\h)$,
and the real function $S(t,\h)$ and the $2n$-dimensional vector function
$Z(t,\h)=(\v P(t,\h),\v X(t,\h))$, which characterize the class
$\CP_\h^t(Z(t,\h),S(t,\h))$, depend regularly on $\sqrt\h$ in the
neighborhood of $\h=0$ and {\em are to be determined}. In the cases where
this does not give rise to ambiguity, we use a shorthand symbol of
$\CP_\h^t$ for $\CP_\h^t(Z(t,\h),S(t,\h))$.

The functions of the class $\CP_\h^t$ are normalized to
\[ \|\Phi (t)\|^2={\lan \Phi (t) |\Phi (t) \ran} \]
in the space $L_2({\Bbb R}^n_x)$ with the scalar product
\[ \lan\Psi (t)|\Phi (t) \ran=
\inl_{\BR^n} d \v x {\Psi}^* (\v x,t,\h)\Phi (\v x,t,\h). \]

In the subsequent manipulation, the argument $t$ in the expression
for the norm may be omitted: $\|\Phi (t)\|^2$= $\|\Phi \|^2$.

In constructing asymptotical solutions, it is useful to define, along with
the class of functions $\CP_\h^t(Z(t,\h),S(t,\h))$, the following class of
functions
\begin{eqnarray}
\lefteqn{\CC_\h^t\big(Z(t,\h),S(t,\h)\big) =}\cr
&\displaystyle=\biggl\{\Phi :\Phi (\v x,t,\h)=
\va\Bigl(\frac{\De\v x}{\sqrt{\h}},t\Bigr)
\exp\Bigl[{\frac{i}{\h}(S(t,\h)+
\lan\v P (t,\h),\De\v x \ran )}\Bigr]\biggr\}, \label{bbst1.6}
\end{eqnarray}
where the functions $\va$, as distinct from (\ref{bbst1.5}), are
independent of $\h$.

At any fixed point in time $t\in{\Bbb R}^1$, the functions belonging to the
class $\CP_\h^t$ {\em are concentrated}, in the limit of $\h\to0$, in the
neighborhood of a point lying on the phase curve $z=Z(t,0)$,
$t\in{\Bbb R}^1$ (the sense of this property is established exactly in
theorems \ref{teor1}--\ref{teor3}\ below). Therefore, it is natural to refer
to the functions of the class $\CP_\h^t$ as {\em trajectory-concentrated
functions}. The definition of the class of trajectory-concentrated functions
includes the phase trajectory $Z(t,\h)$ and the scalar function $S(t,\h)$
as free ``parameters''. It appears that these ``parameters'' are determined
unambiguously from the Hamilton--Ehrenfest equations (see Sec. 3) fitting
the nonlinear ($\vk\ne0$) Hamiltonian of equation (\ref{bbst1.1}). Note
that for a linear Schr\"odinger equation, in the limiting case of $\vk=0$,
the principal term of the series in $\h\to0$ determines the phase
trajectory of the Hamilton system with the Hamiltonian $\CH(\v p,\v
x,t)$, and the function $S(t,0)$ is the classical action along this
trajectory. In particular, in this case, the class $\CP_\h^t$ includes
the well-known dynamic (compressed) coherent states of quantum systems
with quadric Hamiltonians when the amplitude of $\va$ in
(\ref{bbst1.5}) is taken as a Gaussian exponential:
\[ \va(\vec\xi,t)=\exp\Bigl[\frac i2\lan\vec\xi,Q(t)\vec\xi\ran
\Bigr]f(t), \]
where $Q(t)$ is a complex symmetrical matrix with a positive imaginary
part, and the time factor is given by
\[ f(t)=\sqrt[4]{\Im\,Q(t)}\exp\Bigl[-\frac i2\inl_0^t\Im\,Q(\tau)\,d\tau
\Bigr] \]
(see for details \cite{Bagre}).

Let us consider the principal properties of the functions of
the class $\CP_\h^t(Z(t,\h), S(t,\h))$, which are also valid for those
of the class $\CC_\h^t(Z(t,\h),S(t,\h))$.

\begin{teo} \label{teor1}
For the functions of the class $\CP_\h^t(Z(t,\h),S(t,\h))$, the following
asymptotical estimates are valid for centered moments $\De_\al(t,\h)$ of
order $|\al|$, $\al\in \Bbb Z_+^{2n}$:
\begin{equation}
\De_\alpha(t,\h)=\frac{\lan\Phi|\{\De{\hat z}\}^\alpha|\Phi \ran}
{\|\Phi\|^2}=O\big(\h^{|\alpha|/2}\big),\quad\h \to 0.\label{bbst1.8}
\end{equation}
Here, $\{\De{\hat z}\}^\alpha$ denotes the operator with the Weyl symbol
$(\De z)^\alpha$,
\[\De z=z-Z(t,\h)=(\De\v p,\De\v x),\qquad \De\v p=\v p-\v P(t,\h),
\qquad \De\v x=\v x-\v X(t,\h) .\]
\end{teo}
{\bf Proof.} The operator symbol $\{\De\hat z\}^\alpha$ can be written as
\[(\De z)^\alpha=(\De\v p)^{\alpha_p}(\De\v x)^{\alpha_x}, \quad
(\alpha_p,\alpha_x)=\alpha,\]
and, hence, according to the definition of Weyl-ordered
pseudodifferential operators (\ref{bbst1.4}), we have for the mean value
$\sigma_\alpha(t,\h)$ of the operator $\{\De\hat z\}^\alpha$:
\begin{eqnarray*}
\sigma_\alpha(t,\h) &=&
\big\lan \Phi |\{\De{\hat z}\}^\alpha |
\Phi \big\ran=\frac 1 {{(2\pi\h)}^n} \int\limits_{\BR^{3n}} \,d\vec x
\,d\vec y \,d\vec p {\Phi}^*(\vec x,t,\h)\times\\
&\times&\exp \Bigl(\frac i\h\lan(\vec x -\vec y),\vec p\ran\Bigr)
[\De\vec p]^{\alpha_p}\biggl(\frac{\De \vec x+
\De\vec y}{2}\biggr)^{\alpha_x} \Phi(\vec y,t,\h).
\end{eqnarray*}
Here, we have denoted
\[ \De \vec{y}=\vec y -\vec X(t,\h). \]
After the change of variables
\[ \De \vec{x} =\sqrt{\h} \vec\xi,\quad
\De \vec{y} =\sqrt{\h} \vec\zeta, \quad
\De \vec{p} =\sqrt{\h} \vec\omega \]
and taking into consideration the implicit form of the functions
\begin{equation}
\Phi (\vec x,t,\h)=\exp \{i/{\h}(S(t,\h)+\lan \vec P(t,\h),
\De\vec{x}\ran)\}\va\bigg(\frac {\De \vec x} {\sqrt{\h}},t,\h\bigg),
\label{expl}
\end{equation}
belonging to the class $\CP^t_\h(Z(t,\h),S(t,\h))$, we find
\begin{eqnarray*}
\sigma_\alpha(t,\h)&=&
\frac 1 {{(2\pi\h)}^n}{\h}^{{3n}/2}{\h}^{{|\alpha|}/2}2^{-|\alpha_p|}
\int\limits_{\BR^{3n}} d\vec\xi d\vec\zeta d\vec\omega\va^*(\vec\xi, t,\h)
\times\\
&\times& \exp\{i \lan(\vec\xi -\vec\zeta),\vec\omega\ran\}\vec\omega\,
{}^{\alpha_x}(\vec\xi+\vec\zeta)^{\alpha_p}\va(\vec\zeta,t,\h)=\\[8pt]
& =&\h^{(n+|\alpha|)/2}M_\alpha(t,\h),\\
{}\|\Phi\|^2&=&{\h}^{n/2} \int\limits_{\BR^n} d \vec \xi\va^*
(\vec \xi ,t,\h) \va (\vec \xi,t,\h)=\h^{n/2}M_0(t,\h).
\end{eqnarray*}
Since $\va(\vec\xi ,t,\h)$ depends on $\sqrt\h$ regularly and $M_0(t,\h)>0$,
we get
\begin{eqnarray*}
& \De_\alpha(t,\h)=\dac{\sigma_\alpha(t,\h)}{\|\Phi\|^2}
=\h^{|\alpha|/2}\frac{M_\alpha(t,\h)}{M_0(t,\h)}\leqslant\\
&\leqslant\displaystyle\h^{|\alpha|/2}\max_{t \in {[0,T]}}
\frac{M_{\alpha}(t,\h)}{M_0(t,\h)}=O\big(\h^{|\alpha|/2}\big),
\end{eqnarray*}
and thus the theorem is proved.

Let us denote by the symbol $\hat O(\h^\nu)$ an operator $\hat F$, such
that for any function $\Phi$ belonging to the space
$\CP_\h^t(z(t,\h),S(t,\h))$ the asymptotical estimate
\[ \frac{\|\hat F\Phi\|}{\|\Phi\|}=O(\h^\nu),\qquad\h\to 0,  \]
is valid.

\begin{teo} \label{teor2}
For the functions belonging to $\CP_\h^t(Z(t,\h),S(t,\h))$, the following
asymptotical estimates are valid:
\begin{equation}
\{\De{\hat z}\}^\alpha =\hat O({\h}^{|\alpha|/2}),\quad
\alpha\in{\Bbb Z}^{2n}_+,\quad\h \to 0.\label{bbst1.9}
\end{equation}
\end{teo}
{\bf Proof} is similar to that of relation (\ref{bbst1.8}).

\begin{sled} \label{teor2a}
For the functions belonging to $\CP_\h^t(Z(t,\h),S(t,\h))$, the following
asymptotical estimates are valid:
\begin{eqnarray}
& \{ -i\h\pa_t -\dot S(t,\h)+\lan \v P(t,\h), \dot {\v X}(t,\h)\ran +
\lan \dot Z(t,\h),J\De{\hat z}\ran \}=\hat O({\h}),\label{bbst1.9a}\\
& \De{\hat x}_k =\hat O(\sqrt{\h}),\quad
\De{\hat p}_j =\hat O(\sqrt{\h}),\quad k,j=\overline{1,n}.\label{bbst1.10}
\end{eqnarray}
\end{sled}
{\bf Proof} follows from the explicit form (\ref{expl}) of the
trajectory-concentrated functions [$\Phi(\vec x,t,\h)\in\CP_\h^t$
(\ref{bbst1.6})] and from the estimates (\ref{bbst1.9}).

\begin{teo} \label{teor3}
For any function $\Phi(\vec x,t,\h)\in\CP_\h^t(Z(t,\h),S(t,\h))$, the
limiting relations
\begin{eqnarray}
&&\lim _{\h \to 0} \frac 1{\| \Phi\|^2} |\Phi(\vec x,t,\h)|^2=
\de (\vec x-\vec X(t,0)),\label{bbst1.11}\\
&&\lim_{\h\to 0} \frac 1{\| \tilde\Phi\|^2} |\tilde\Phi(\vec p,t,\h)|^2=
\de (\vec p-\vec P(t,0)),\label{bbst1.12}
\end{eqnarray}
where $\tilde\Phi(\vec p,t,\h)=F_{\h,\vec x\to\vec p}\Phi(\vec x,t,\h)$,
$F_{\h,\vec x\to\vec p}$ is the $\h^{-1}$ Fourier transform {\rm\cite{Mas4}},
are valid.
\end{teo}
{\bf Proof.} Let us consider an arbitrary function $\phi(\vec{x})\in
{\Bbb S}$. Then for any function $\Phi(\vec x,t,\h)\in\CP_\h^t$ the integral
\[ \Big\lan\frac{|\Phi(t,\h)|^2}{\|\Phi(t,\h)\|^2}\Big|\,\phi\Big\ran=
\frac 1{\|\Phi(t,\h)\|^2}\inl_{\BR^n_x} \phi(\vec x)|
\Phi(\vec x,t,\h)|^2\,d\vec x=\frac 1{\|\va(t,\h)\|^2}\inl_{\BR^n_x}
\phi(\vec x)\Big|\va\Bigl(\frac{\De\vec x}{\sqrt\h},t\Bigr)\Big|^2\,d\vec x \]
after the change of variables $\vec\xi=\De\vec x/\sqrt\h$, becomes
\[ \big\lan|\Phi(t,\h)|^2\big|\,\phi\big\ran=\frac{\h^{n/2}}{\|\va(t,\h)\|^2}
\inl_{\BR^n_\xi} \phi(\vec X(t,\h)+\sqrt\h\vec\xi)|\va(\vec\xi,t,\h)|^2\,
d\vec\xi. \]
Let us pass in the last equality to the limit of $\h\to0$, and, in view of
\[ \|\va(t,\h)\|^2=\h^{n/2}\inl_{\BR^n_\xi} |\va(\vec\xi,t,\h)|^2\,
d\vec\xi, \]
and a regular dependence of the function $\va(\vec\xi,t,\h)$ on
$\sqrt\h$, we arrive at the required statement.

The proof of relation (\ref{bbst1.12}) is similar to the previous one if
we notice that the Fourier transform of the function
$\Phi(\vec x,t,\h)\in\CP_\h^t$ can be represented as
\[ \tilde\Phi(\vec p,t,\h)=\exp\Bigl\{\frac i\h\big[S(t,\h)-
\lan\vec p,\vec X(t,\h)\ran\big]\Bigr\}\tilde\va\Bigl(
\frac{\vec p-\vec P(t,\h)}{\sqrt\h},t,\h\Bigr), \]
where
\[ \tilde\va(\vec\omega,t,\h)=\frac 1{(2\pi)^{n/2}}\inl_{\BR^n_\xi}
e^{-i\lan\vec\omega,\vec\xi\ran}\va(\vec\xi,t,\h)d\xi. \]

Denote by $\lan\hat L(t)\ran$ the mean value of the operator
$\hat L(t)$, $t\in\BR^1$, self-conjugate in $L_2(\BR^n_x)$, calculated
from the function $\Phi(\vec x,t,\h)\in\CP_\h^t$. Then the
following corollary is valid:

\begin{sled} \label{teor4}
For any function $\Phi(\vec x,t,\h)\in\CP_\h^t(Z(t,\h),S(t,\h))$ and any
operator $\hat A(t,\h)$ whose Weyl symbol $A(z, t,\h)$ satisfies
Supposition \ref{assum1}, the equality
\begin{eqnarray}
&\displaystyle\lim _{\h \to 0} \lan\hat A(t,\h)\ran=\lim _{\h \to 0}
\frac 1{\|\Phi\|^2} \lan\Phi(\vec x,t,\h)|\hat A(t,\h)|\Phi(\vec
x,t,\h)\ran =\vspace{12pt}\cr
&=A(Z(t,0),t,0)\label{bbst1.13}
\end{eqnarray}
is valid.
\end{sled}
{\bf Proof} is similar to that of relations (\ref{bbst1.11}) and
(\ref{bbst1.12}).

Following \cite{Bagre}, we introduce

\begin{defin}\label{def1}  We refer to the solution
$\Phi(\vec x,t,\h)\in \CP_\h^t$ of equation $($\ref{bbst1.1}$)$
as quasi-classically concentrated on the phase trajectory $Z(t,\h)$ for
$\h\to0$, provided that the conditions $($\ref{bbst1.11}$)$ and
$($\ref{bbst1.12}$)$ are fulfilled.
\end{defin}

\begin{rem} The above estimates {\rm(\ref{bbst1.9})} of operators
$\{\De\hat z\}^\alpha$ allow a consistent expansion of the functions of
the class $\CP_\h^t(Z(t,\h),S(t,\h))$ and the operator of equation
{\rm(\ref{bbst1.1})} in a power series for $\sqrt\h$. This expansion
gives rise to a set of recurrent equations which determine the
sought-for asymptotical solution of equation {\rm(\ref{bbst1.1})}.

For any function $\Phi\in\CP_\h^t(Z(t,\h),S(t,\h))$, the representation
\begin{equation}\label{funcexp}
\Phi (\vec x, t,\h)=
\sum_{k=0}^N\h^{k/2}\Phi^{(k)}(\v x,t,\h)+ O(\h^{(N+1)/2}),
\end{equation}
where $\Phi^{(k)}(\v x,t,\h)\in\CC_\h^t(Z(t,\h),S(t,\h))$, is valid.
Representation {\rm(\ref{funcexp})} naturally induces the
expansion of the space $\CP_\h^t(Z(t,\h),S(t,\h))$ in a direct sum of
subspaces
\begin{equation}  \label{expan}
\CP_\h^t(Z(t,\h),S(t,\h))=
\mathop{\oplus}\limits_{l=0}^{\infty}\CP_\h^t(Z(t,\h),S(t,\h),l).
\end{equation}
Here, the functions $\Phi \in \CP_\h^t(Z(t,\h),S(t,\h),l)\subset
\CP_\h^t(Z(t,\h),S(t,\h))$, according to {\rm(\ref{bbst1.6})}, have
estimates by the norm
\begin{equation}\label{norm}
\displaystyle \frac{1}{\h^{n/2}}\|\Phi\|_{L _2(\BR^n_x)}=\h^{l/2}\mu (t),
\end{equation}
where the function $\mu(t)$ is independent of $\h$ and continuously
differentiable with respect to $t$.

Similar to the proof of the estimates {\rm(\ref{bbst1.9}) and
(\ref{bbst1.9a})}, it can be shown that the operators
\[\{\De{\hat z}\}^\alpha, \qquad \{ -i\h\pa_t -\dot S(t,\h)+\lan \v P(t,\h),
\dot{\v X}(t,\h)\ran + \lan \dot Z(t,\h), J\De{\hat z}\ran\} \]
do not disrupt the structure of the expansion {\rm(\ref{funcexp}),
(\ref{expan}),} and
\begin{equation} \label{struct}
\begin{array}{l}
\{\De{\hat z}\}^\alpha: \CP_\h^t(Z(t,\h),S(t,\h),l)
\rightarrow   \CP_\h^t(Z(t,\h),S(t,\h),l+|\alpha |),\\[8pt]
\{ -i\h\pa_t -\dot S(t,\h)+\lan \v P(t,\h), \dot {\v X}(t,\h)\ran +
\lan \dot Z(t,\h), J\De{\hat z}\ran \}:\\[8pt]
\qquad :\CP_\h^t(Z(t,\h),S(t,\h),l)
\rightarrow   \CP_\h^t(Z(t,\h),S(t,\h),l+2).
\end{array}
\end{equation}
\end{rem}

\begin{rem}  From Corollary {\rm\ref{teor4}} it follows that the
solution $\Psi(\vec x,t,\h)$ of equation {\rm(\ref{bbst1.1}),}
belonging to the class $\CP_\h^t,$, is quasi-classically concentrated.
\end{rem}

The limiting character of the conditions (\ref{bbst1.11}) and (\ref{bbst1.12})
and the asymptotical character of the estimates
(\ref{bbst1.8})--(\ref{bbst1.9a}) valid for the class of
trajectory-concentrated functions make it possible to construct
quasi-classically concentrated solutions to the Hartree type equation
{\em not exactly}, but {\em approximately}. In this case, the $L_2$ norm
of the error has an order of $\h^\alpha$, $\alpha>1$ for $\h\to0$ on any
finite time interval $[0,T]$. Denote such an approximate solution as
$\Psi_{\rm as}=\Psi_{\rm as}(\vec x,t,\h)$.  This solution satisfies the
following problem:
\begin{eqnarray}
& \Bigl[-i\h\dac\pa{\pa t}+\hat\CH(t)+\vk\hat V(t,\Psi_{\rm as})\Bigr]
\Psi_{\rm as}=O(\h^\alpha),\label{zadacha1}\\
& \Psi_{\rm as}\in\CP_\h^t(Z(t,\h),S(t,\h),\h), \quad
t\in[0,T],\label{zadacha2}
\end{eqnarray}
where $O(\h^\alpha)$ denotes the function $g^{(\alpha)}(\vec x,t,\h)$,
the ``{\em residual}'' of equation (\ref{bbst1.1}). For the residual,
the following estimate is valid:
\begin{equation} \max_{0\le t\le T} \|g^{(\alpha)}(\vec x,t,\h)\|=
O(\h^\alpha), \qquad \h\to0.\label{nevyazka}
\end{equation}
Below we refer to the function $\Psi_{\rm as}(\vec x,t,\h)$ satisfying
the problem (\ref{zadacha1})--(\ref{nevyazka}) as a {\em quasi-classically
concentrated solution} ($\bmod\h^\alpha$, $\h\to0$) of the Hartree type
equation (\ref{bbst1.1}).

The main goal of this work is to construct quasi-classically
concentrated solutions to the Hartree type equation (\ref{bbst1.1}) with any
degree of accuracy in small parameter $\sqrt\h$, $\h\to0$, i.e.,
functions $\Psi_{\rm as}(\vec x,t,\h)=\Psi^{(N)}(\vec x,t,\h)$
satisfying the problem (\ref{zadacha1})--(\ref{nevyazka}) in
$\bmod(\h^{(N+1)/2})$, where $N\ge2$ is any natural number.

Thus, the quasi-classically concentrated solutions $\Psi^{(N)}(\vec x,t,\h)$
of the Hartree type equation describe approximately the evolution of the
initial state $\Psi_0(\vec x,\h)$ if the latter has been taken from a class
of trajectory-concentrated functions $\CP_\h^0$. The operators $\hat\CH(t)$
and $\hat V(t,\Psi)$ entering in the Hartree type equation
(\ref{bbst1.1}) leave the class $\CP_\h^t$ invariant on a finite time
interval $0\le t\le T$ since their symbols satisfy Supposition
\ref{assum1}.  Therefore, in  constructing quasi-classically
concentrated solutions to the Cauchy problem, the initial conditions
can be taken in the form
\begin{equation} \Psi(\v x,t,\h)|_{t=0}=\Psi_0(\v x,\h), \qquad
\Psi_0\in\CP^0_\h(z_0,S_0).\label{bbst1.7}
\end{equation}
The functions from the class $\CP_\h^0$ have the following form:
\begin{equation} \Psi_0(\vec x,\h)=\exp\Bigl\{\frac i\h[S(0,\h)+
\lan\vec P_0(\h), (\vec x-\vec X_0(\h))\ran]\Bigr\}\va_0
\Bigl(\frac{\vec x-\vec X_0(\h)}{\sqrt\h},\h\Bigr),\quad
\va_0(\vec\xi,\h)\in\BS(\BR^n_\xi),\label{vid}
\end{equation}
where $Z_0(\h)=(\vec P_0(\h),\vec X_0(\h))$ is an arbitrary point of the
phase space $\BR^{2n}_{px}$, and the constant $S_0(\h)$ can be put equal
to zero without loss of generality.

Important particular cases of the initial conditions of type
(\ref{vid}) are
$$ \va_0(\vec\xi)=e^{-\lan\vec\xi,A\vec\xi\ran/2},
\leqno{1)}$$
where the real $n\times n$ matrix $A$ is positive definite and symmetrical.
Then relationship (\ref{vid}) defines the {\em Gaussian packet};
$$ \va_0(\vec\xi)=e^{i\lan\vec\xi,Q\vec\xi\ran/2}H_\nu(\Im\,Q\vec\xi),
\leqno{2)}$$
where the complex $n\times n$ matrix $Q$ is symmetrical and has a positive
definite imaginary part $\Im\,Q$ and $H_\nu(\vec\eta)$ and $\vec\eta\in\BR^n$
are multidimensional Hermite polynomials of multi-index
$\nu=(\nu_1,\dots,\nu_n)$ \cite{Beitman2}. In this case, relationship
(\ref{vid}) defines the {\em Fock states of a multidimensional oscillator}.

The solution of the Cauchy problem (\ref{bbst1.1}), (\ref{bbst1.7})
leads in turn to a set of Hamilton--Ehrenfest equations to the study of
which we are coming.

\section{The set of Hamilton--Ehrenfest equations}

In view of Supposition \ref{assum1}\ for the symbols $\CH(z,t)$ and $V(z,w,t)$,
the operator $\CH (\hat z,t)$ (\ref{bbst1.2}) is self-conjugate to the scalar
product $\lan\Psi |\Phi\ran$ in the space $L_2(\BR_x^n)$ and
the operator $V(\hat z,\hat w ,t)$ (\ref{bbst1.3}) is self-conjugate to the
scalar product $L_2(\BR^{2n}_{xy})$:
\[ \lan\Psi(t)|\Phi(t)\ran_{\BR^{2n}}
=\dil_{\BR^{2n}} d\v xd\v y {\Psi}^* (\v x,\v y,t,\h)
\Phi(\v x,\v y,t,\h). \]

Therefore, for the exact solutions of equation (\ref{bbst1.1}) we have
\[ \|\Psi(t)\|^2 =\|\Psi(0)\|^2, \]
and for the mean values of the operator $\hat A(t)=A(\hat z,t)$,
calculated for these solutions, the equality
\begin{eqnarray}
& \dac{d}{dt}\lan\hat A(t)\ran =\Bigl\lan\frac{\pa\hat A(t)}
{\pa t}\Bigr\ran +\frac{i}{\h}\lan[\hat {\CH},\hat A (t)]_-\ran +\cr
& +\dac{i\vk}{\h}\Bigl\lan\int d\v y\,\Psi^*(\v y,t,\h)
  [\hat A (t),V(\hat z,\hat w,t)]_-\Psi(\v y,t,\h)\Bigl\ran,
\label{bbst2.1}
\end{eqnarray}
where $[\hat A,\hat B]_-=\hat A\hat B-\hat B\hat A$ is the commutator of
the operators $\hat A$ and $\hat B$, is valid.  We refer to
equation (\ref{bbst2.1}) as the {\em Ehrenfest equation for the operator
$\hat A$ and function}\ $\Psi(\v x,t,\h)$. This term was chosen in view
of the fact that in the linear case ($\vk=0$) equation
(\ref{bbst1.1}) goes into the quantum mechanical Schr\"odinger equation
and relationship (\ref{bbst2.1}) into the Ehrenfest equation
\cite{Ehrenfest}.

Let us make the following notations:
\begin{equation}
  \hat z =(\hat {\v p},\hat {\v x}),\quad Z(t,\h)=(\vec P(t,\h),
  \vec X(t,\h)),\quad\Delta\hat z =\hat z -Z(t,\h).
  \label{bbst2.2}
\end{equation}
Using the rules of composition for Weyl symbols \cite{Karas2}, we find
for the symbol of the operator $\hat C=$ $\hat A\,\hat B$
\begin{equation}
C(z)=A\Big(\stackrel{2}{z}+\frac{i\h}2J\frac{\stackrel{1}{\pa}}
{\pa z}\Big) B(z)=B \Big(\stackrel{2}{z}-\dac{i\h}2J
\frac{\stackrel{1}{\pa}}{\pa z}\Big) A( z).
\end{equation}
Here, the index over an operator symbol specifies the turn of its
action.  We suppose that for the Hartree type equation (\ref{bbst1.1}),
exact solutions [or solutions differing from exact ones by a quantity
$O(\h^\iy)$] exist in the class of trajectory-concentrated functions.
Let us write Ehrenfest equations (\ref{bbst2.1}) for the mean
values of the operators $\hat z_j$,$\{\Delta\hat z\}^\alpha$
calculated from such (trajectory-coherent) solutions of equation
(\ref{bbst1.1}).  After cumbersome, but not complicated calculations
similar to the calculations that were performed for the linear case with
$\vk=0$ (see for details \cite{Bagre}), we then obtain, restricting
ourselves to the moments of order $N$, the following set of ordinary
differential equations:
\begin{eqnarray}
&\displaystyle\dot z=\sum_{|\mu|=0}^N \frac 1{\mu!}
J\Big(\CH_{z\mu}(z,t)\De_\mu +\tvk\sum_{|\nu|=0}^N\frac 1{\nu!}
 V_{z\mu\nu}(z,t)\De_{\nu}\Big),\cr
&\dot\De_{\alpha}=\dum_{|\mu+\gamma|=0}^N \Big(-i\h\Big)^{|\gamma|-1}\,
\frac{[(-1)^{|\gamma_p|}-(-1)^{|\gamma_x|}]\alpha!\beta!
\theta(\alpha-\gamma)\theta(\beta-\gamma)}
{\gamma!(\alpha-\gamma)!(\beta-\gamma)!\mu!}\times \label{bbst2.4}\\
&\times\displaystyle
\Big(\CH_\mu(z,t)+\tvk\sum_{|\nu|=0}^N\frac 1{\nu!}
 V_{\mu\nu}(z,t)\De_{\nu}\Big)\De_{\alpha-\gamma +J\beta-J\gamma}-
\sum_{k=1}^{2n}\dot z_k \alpha_k \De_{\alpha(k)}\nonumber
\end{eqnarray}
with initial conditions
\begin{eqnarray}
& z\big|_{t=0}=z_0=\lan\Psi_0|\hat z|\Psi_0\ran, \qquad
\De_\alpha\big|_{t=0}=\lan\Psi_0|\{\hat z-z_0\}^\alpha|\Psi_0\ran,
\label{nul}\\
&\alpha\in{\Bbb Z}^{2n}_+, \qquad |\alpha|\le N.\nonumber
\end{eqnarray}
Here, $\tvk=\vk\|\Psi_0(\v x,\h)\|^2$ and $\Psi_0(\v x,\h)$ is the initial
function from (\ref{bbst1.7}),
\begin{eqnarray}
& \CH_\mu(z,t)=\dac{\pa^{|\mu|}\CH ( z,t)}{\pa z^\mu},\quad
V_{\mu\nu}(z,t)= \frac{\pa^{|\mu+\nu|} V( z, w,t)}
 {\pa z^\mu \pa w^\nu}\Big|_{\omega=z}\vspace{4pt} \cr
&\CH_{z\mu}(t,\h)= \pa_z{\CH}_\mu(t,\h),\quad
\alpha=(\alpha_p,\alpha_x), \quad J\alpha=(\alpha_x,\alpha_p) ,
\label{bbst2.5}\\[2pt]
&\theta(\alpha-\beta)=\prod\limits_{k=1}^{2n}\theta(\alpha_k-\beta_k),\quad
\alpha(k)=(\alpha_1-\de_{1,k}, \dots, \alpha_{2n}-\de_{2n,k}).\nonumber
\end{eqnarray}

By analogy with the linear theory ($\vk=0$) \cite{Bagre}, we refer to
equations (\ref{bbst2.4}) as {\em Hamilton--Ehrenfest equations} of order
$N$. In view of the estimates (\ref{bbst1.8}), these equations are
equivalent, for the class $\CP_\h^t$, to the nonlinear Hartree type
equation (\ref{bbst1.1}) accurate to $O(\h^{(N+1)/2})$.

For the case of $N=2$, the Hamilton--Ehrenfest equations take the form
\begin{equation}
\left\{\begin{array}{l}
 \displaystyle\dot z=J\pa_z\Bigl(1+\frac 12\lan\pa_z,
 \De_2\pa_z\ran+\frac 12\lan\pa_\omega,
 \De_2\pa_\omega\ran\Bigl)(\CH(z,t)+\\[8pt]
 \qquad\displaystyle +\tvk V(z,\omega,t))|_{\omega=z},\\[8pt]
 \dot\De_2=JM\De_2-\De_2 MJ,\end{array}\right.\label{bbst2.6}
\end{equation}
where
$$
 M=\bigl[\CH_{zz}(z,t)+\tvk V_{zz}(z,\omega,t)\bigr]
\Bigl|_{\omega=z}.
$$

Equations (\ref{bbst2.6}) can be written in the equivalent form if in the
second equation we put
\[\De_2(t)= A(t)\De_2(0) A^+(t), \]
and then it becomes
\begin{equation}
\dot A=JMA \qquad A(0)=\Bbb I.  \label{bbst2.6a}
\end{equation}

\section{Linearization of the Hartree type equation}

Let us now construct a quasi-classically concentrated (for $\h\to0$) solution
to equation (\ref{bbst1.1}), satisfying the initial condition
(\ref{bbst1.7}).

Designate by
\begin{equation}
y^{(N)}(t,\h)=(Z_{j_1},\De^{(2)}_{j_2j_3}, \De^{(3)}_{j_4j_5j_6},\dots)
=(Z(t,\h),\De_\alpha(t,\h)), \quad |\alpha|\le N
\label{bbst3.1}
\end{equation}
the solution of the Hamilton--Ehrenfest equations of
order $N$ (\ref{bbst2.4}) with the initial data $y^{(N)}(0,\h)$
(\ref{nul}) determined by the initial function $\Psi_0(\v x,\h)$
(\ref{bbst1.7}), i.e., the mean values $Z(0,\h)$ and $\De_\al(0,\h)$
are calculated from the function $\Psi_0(\v x,\h)$.  Let us expand the
``kernel'' of the operator $\hat V(t,\Psi)$ in a Taylor power series
for the operators $\De\hat w=\hat w-Z(t,\h)$:
\begin{equation}
V(\hat z,\hat w,t)=\sum_{|\alpha|=0}^\iy\frac 1{\alpha!}
\frac{\pa^{|\alpha|}V(\hat z,w,t)}{\pa w^\alpha}\Big|_{w=Z(t,\h)}
\{\De\hat w\}^\alpha. \label{nucl}
\end{equation}
Substituting this series into equation (\ref{bbst1.1}), we obtain for
the functions $\Psi\in\CP_\h^t$
\begin{eqnarray}
&\Bigl[-i\h\pa_t+\CH(\hat z,t)+\tvk \dum_{|\alpha|=0}^N\frac 1{\alpha!}
\frac{\pa^{|\alpha|}V(\hat z, w,t)}{\pa w^\alpha}\Big|_{w=Z(t,\h)}
\De_\alpha(t,\h)\Bigr]\Psi=\cr
& =O(\h^{(N+1)/2}), \qquad \Psi\big|_{t=0}=\Psi_0;.\label{bbst3.2}
\end{eqnarray}
where
\begin{equation}
\begin{array}{l}
Z(t,\h)=\dac 1{\|\Psi(t,\h)\|^2}\lan\Psi(t,\h)|\hat
z|\Psi(t,\h)\ran;\\[8pt] \De_\alpha(t,\h)=\dac
1{\|\Psi(t,\h)\|^2}\lan\Psi(t,\h)| \{\De\hat
z\}^\alpha|\Psi(t,\h)\ran.\end{array}\label{nucl1} \end{equation} In
view of the asymptotical estimates (\ref{bbst1.8}), the functions
$z(t,\h)$ and $\De_\alpha(t,\h)$ can be determined with any degree of
accuracy from the Hamilton--Ehrenfest equations (\ref{bbst2.4}) as
\begin{equation}
\begin{array}{l}
z(t,\h)=z(t,\h,N)+O(\h^{(N+1)/2});\\[8pt]
\De_\alpha(t,\h)=\De_\alpha(t,\h,N)+O(\h^{(N+1)/2}) \quad
|\alpha|\le N,\end{array}\label{nucl2}
\end{equation}
where $z(t,\h,N)$ and $\De_\alpha(t,\h,N)$ are solutions of the
Hamilton--Ehrenfest equations of order $N$, which are completely determined
by the initial condition of the Cauchy problem for the Hartree type equation,
$\Psi_0(\v x,t,\h)$, and do not use the explicit form of the solution
$\Psi(\v x,t,\h)$ (\ref{bbst3.2}).  Thus, the change of the mean values
of the operators for the solutions of the Hamilton--Ehrenfest equations
of order $N$ (\ref{nucl2}) {\em linearizes} the Hartree type
equation (\ref{bbst3.2}) accurate to $O(\h^{(N+1)/2})$.  So, to find an
asymptotical solution to the Hartree type equation (\ref{bbst1.1}), we
should consider the {\em linear} Schr\"odinger type equation:
\begin{eqnarray}
&\hat L^{(N)}(t,\Psi_0)\Phi=O(\h^{(N+1)/2}), \quad
\Phi\big|_{t=0}=\Phi_0;\label{bbst3.2b}\\
&\hat L^{(N)}(t,\Psi_0)=-i\h\pa_t + \CH (\hat z,t)
+\tvk \displaystyle\sum_{|\alpha|=0}^N\frac 1{\alpha!}
\frac{\pa^{|\alpha|}V(\hat z, w,t)}{\pa w^\alpha}\Big|_{w=Z(t,\h,N)}
\De_\alpha(t,\h,N).\nonumber
\end{eqnarray}

\begin{defin} We call an equation of type {\rm(\ref{bbst3.2b})} with a given
$\Psi_0$ a Hartree equation in the trajectory-coherent approximation or
a linear associated Schr\"odinger equation of order $N$ for the
Hartree type equation {\rm(\ref{bbst1.1}).}
\end{defin}

The following statement is valid:

\begin{demo} \label{d1} If the function
$\Phi^{(N)}(\v x,t,\h,\Psi_0)\in\CP_\h^t$ is an asymptotical $($accurate to
$O(\h^{(N+1)/2})$, $\h\to0)$ solution of equation {\rm(\ref{bbst3.2b}),}
satisfying the initial condition $\Phi|_{t=0}=\Psi_0$, the function
\[ \Psi^{(N)}(\v x,t,\h)=\Phi^{(N)}(\v x,t,\h,\Psi_0) \]
is an asymptotical $($accurate to $O(\h^{(N+1)/2})$, $\h\to0)$ solution
of the Hartree type equation {\rm(\ref{bbst1.1}).}
\end{demo}

Now we expand the operators
\[\CH(\hat z,t),\quad  \frac{\pa^{|\alpha|}V
(\hat z, w, t)}{\pa w^\alpha}\Big|_{w=Z(t,\h,N)}\]
in a Taylor power series for the operator $\De\hat z$ and present the
operator $-i\h\pa_t$ in the form
\begin{eqnarray*}
&-i\h\pa_t = \{-\lan \v P(t,\h,N),\dot {\v X}(t,\h,N)\ran+\dot S(t,\h)\} -
\lan \dot Z(t,\h,N),J\De{\hat z}\ran+\\
&+ \{ -i\h\pa_t -\dot S(t,\h)+\lan \v
P(t,\h,N), \dot {\v X}(t,\h,N)\ran + \lan \dot Z(t,\h,N),J\De{\hat z}\ran\}.
\end{eqnarray*}
Here, the group of terms in braces containing $-i\h\pa_t$, in
view of (\ref{struct}), has an order of $\hat O(\h)$. Other terms can be
estimated, in view of (\ref{bbst1.9a}), by the parameter $\h$. Substitute
the obtained expansions into (\ref{bbst3.2b}). Take (accurate to
$O(\h^{N/2})$) the real function $S(t,\h)$ entering in the definition of
the class $\CP_\h^t(Z(t,\h),S(t,\h))$ in the form
\begin{eqnarray}
&S(t,\h)=S^{(N)}(t,\h)= \dil_0^t \Bigl\{\lan\vec
P(t,\h,N)\dot{\vec X}(t,\h,N)\ran- \CH(Z(t,\h,N),t)-
\cr &-\tvk\dum_{|\alpha|=0}^N\frac 1{\alpha!} \frac{\pa^{|\alpha|}V(Z(t,\h,N),
w,t)}{\pa w^\alpha}\Big|_{w=Z(t,\h,N)}
\De_\alpha(t,\h,N)\Bigr\}dt.\label{kak15}
\end{eqnarray}
As a result, equation (\ref{bbst3.2b}) will not contain operators
of multiplication by functions depending only on $t$ and $\h$.

In view of the estimates (\ref{bbst1.9}) and (\ref{bbst1.9a}) valid for
the class $\CP_\h^t(Z(t,\h),S(t,\h))$, we obtain for (\ref{bbst3.2})
\begin{equation}
\Big\{ -i\h\pa_t + \hat\FH_0(t,\Psi_0)+\h\hat\FH^{(N)}(t,\Psi_0)\Big\}\Phi
=O(\h^{(N+1)/2}),\label{bbst3.4}
\end{equation}
with the following notations:
\begin{eqnarray}
&&\hat\FH^{(N)}(t,\Psi_0)=\sum_{k=1}^N\h^{k/2}\hat\FH_k(t,\Psi_0),
\label{bbst3.5a} \\
&&\hat\FH_0(t,\Psi_0)=-\dot S(t,\h)+
\lan \v P(t,\h), \dot {\v X}(t,\h)\ran +
\lan \dot Z(t,\h),J\De{\hat z}\ran +
\dac 12 \lan\De\hat z,\FH_{zz}(t,\Psi_0)\De\hat z\ran, \label{bbst3.5} \\
&&\FH_{zz}(t,\Psi_0)=\Bigl[\CH_{zz}(z,t)+\tvk V_{zz}(z,w,t)\Bigr]
\Big|_{z=w=Z(t,\h,N)},\vspace{12pt}\cr
&& \h^{(k+2)/2}\hat\FH_k(t,\Psi_0)=-\lan \dot Z_{(k)}(t),J\De\hat z\ran
 +\dum_{|\alpha|=k+2}\frac 1{\alpha!}\frac{\pa^{|\alpha|}\CH(z,t)}
{\pa z^\alpha}\Big|_{z=Z(t,\h,N)} \{\De\hat z\}^\alpha+\cr
&&\quad +\tvk \dum_{|\alpha+\beta|=k+2}\frac 1
{\alpha!\beta!}\frac{\pa^{|\alpha+\beta|}V(z,w,t)}{\pa
z^\beta\pa w^\alpha}\Big|_{z=w=Z(t,\h,N)}\{\De\hat
z\}^\beta\De_\alpha(t,\h,N).\label{bbst3.6}
\end{eqnarray}
Here, $k=\overline{1,N}$ and the functions $Z_{(k)}(t)$ are the
coefficients of the expansion of the projection $Z(t,\h)$ of the
solution $y^{(N)}(t,\h)$ of the Hamilton--Ehrenfest equations on the
phase space $\BR^{2n}$ in a power series of the regular perturbation
theory for $\sqrt\h$:
\[ Z(t,\h)=Z(t,\h,N)=Z(t,0)+\sum_{k=2}^N \h^{k/2} Z_{(k)}(t). \]
From the Hamilton--Ehrenfest equations, in view of the
fact that the first-order moments are zero ($\De_\alpha(t,\h,N)=0$ for
$|\al|=1)$, it follows that the coefficient $\dot Z_{(1)}(t)$ is equal
to zero.

\begin{rem} The solutions of the set of Hamilton--Ehrenfest equations
depend on the index $N$ that denotes the highest order of the centered
moments $\De_\alpha$, $\alpha\in\BZ^{2n}_+$. We shall omit the index
$N$ if this does not give rise no ambiguity.
\end{rem}

The operators $\hat\FH_0(t)$ (\ref{bbst3.5}) and $\hat\FH_k(t)$
(\ref{bbst3.6}) depend on the mean $Z(t,\h)$ and moments $\De_\al (t,\h)$,
i.e., on the solution $y^{(N)}(t,\h)$ of the Hamilton--Ehrenfest
equations (\ref{bbst2.4}). The solutions of equation (\ref{bbst3.4}) in
turn depend implicitly on $y^{(N)}(t,\h)$:
\[\Phi(\v x,t,\h)=\Phi(\v x,t,\h,y^{(N)}(t,\h)). \]
Below the function arguments $y^{(N)}(t,\h)$
or $\Psi_0$ can be omitted if this does not give rise to ambiguity.
For example, we may put $\FH_0(t)=\hat\FH_0(t,\Psi_0)$.

In accordance with the expansion (\ref{expan}) and (\ref{funcexp}), the
solution of equation (\ref{bbst3.4}) can be represented in the form
\begin{equation}
\Phi(\v x,t,\h,\Psi_0)=
\sum_{k=0}^N\h^{k/2}\Phi^{(k)}(\v x,t,\h,\Psi_0)+
O(\h^{(N+1)/2}),\label{bbst3.7}
\end{equation}
where
\[ \Phi^{(k)}(\v x,t,\h,\Psi_0)\in\CC_\h^t(Z(t,\h),S(t,\h)). \]
In view of (\ref{struct}), for the operators $\{-i\h\pa_t+
\hat\FH_0(t,\Psi_0)\}$ (\ref{bbst3.5}) and
$\h^{(k+2)/2}\hat\FH_k(t,\Psi_0)$, $k=\overline{1,N}$ (\ref{bbst3.6})
the following is valid:
\begin{equation} \label{struct1}
\begin{array}{l}
\h^{(k+2)/2}\hat\FH_k(t,\Psi_0): \CP_\h^t(Z(t,\h),S(t,\h),l)
\rightarrow   \CP_\h^t(Z(t,\h),S(t,\h),l+k+2),\\[8pt]
\{ -i\h\pa_t +\hat\FH_0(t,\Psi_0)\} :\CP_\h^t(Z(t,\h),S(t,\h),l)
\rightarrow   \CP_\h^t(Z(t,\h),S(t,\h),l+2).
\end{array}
\end{equation}
Substitute (\ref{bbst3.7}) into (\ref{bbst3.4}) and equate the terms having
the same order in $\h^{1/2}, \h \to 0$ in the sense of (\ref{struct1}).
As a result we obtain a set of recurrent {\em associated linear equations}
of order $k$ to determine the functions $\Phi^{(k)}(\v x,t,\h,\Psi_0)$:
\begin{eqnarray}
&\h^1&{}\{ -i\h\pa_t+\hat\FH_0(t,\Psi_0)\} \Phi^{(0)}=0,\label{kak5}\\
&\h^{3/2}&{}\{-i\h\pa_t + \hat\FH_0(t,\Psi_0)\} \Phi^{(1)}+
\h\hat\FH_1(t,\Psi_0)\Phi^{(0)}= 0,\label{bbst3.9}\\
&\h^{2}&{}\{ -i\h\pa_t + \hat\FH_0(t,\Psi_0)\} \Phi^{(2)}+
\h\hat\FH_1(t,\Psi_0)\Phi^{(1)} +\h^{3/2}\hat\FH_2(t,\Psi_0)\Phi^{(0)}=0,
\label{bbst3.10}\\
&&\dots\dots\dots\dots\dots\dots\dots\nonumber
\end{eqnarray}

It is natural to call equation (\ref{kak5}) for the principal term of the
asymptotical solution as the Hartree type equation in the
trajectory-coherent approximation in $\bmod\h^{3/2}$. This equation is
the Schr\"odinger equation with the Hamiltonian quadric with respect to
the operators $\hat{\v p}$ and $\hat{\v x}$.

\section{The trajectory-coherent solutions of the Hartree type equation}

The solution of the Schr\"odinger equation with a quadric Hamiltonian
is well known \cite{Chern, Manko}.  For our purposes, it is convenient to
take quasi-classical trajectory-coherent states (TCS's) \cite{Bagre} as a
basis of solutions to equation (\ref{kak5}). We shall refer to the solution
of the nonlinear Hartree type equation, which coincides with the TCS at
the time zero, as a {\em trajectory-coherent solution of the Hartree
type equation}. Now we pass to constructing solutions like this.

Let us write the symmetry operators $\hat a(t,\Psi_0)$ of equation
(\ref{kak5}), linear with respect to the operators $\De\hat z$, in the form
\begin{equation}
\hat a(t,\Psi_0)=N_a\lan b(t,\Psi_0),\De\hat z\ran,\label{kak6}
\end{equation}
where $N_a$ is a constant and $b(t)$ is a $2n$-space vector.  From the
equation
\begin{equation}
-i\h\frac{\pa\hat a(t)}{\pa t}+[\hat\FH_0(t,\Psi_0),\hat
a(t)]_-=0,\label{kak7}
\end{equation}
which determines the operators $\hat a(t)$, in view of the explicit form
of the operator $\hat\FH_0(t,\Psi_0)$ (\ref{bbst3.5}), we obtain
\begin{eqnarray*}
& -i\h\lan\dot b(t),\De\hat z\ran+i\h\lan b(t),\dot Z(t,\h)\ran+\\
& +\Bigl[\Bigl\{-\dot S(t,\h)+
\lan \v P(t,\h), \dot {\v X}(t,\h)\ran +
\lan \dot Z(t,\h),J\De{\hat z}\ran+\dac 12
\lan\De z,\FH_{zz}(t,\Psi_0)\De\hat z\ran\Bigr\},\lan b(t),\De\hat z\ran
\Bigr]=0.
\end{eqnarray*}
Taking into account the commutative relations
\[ [\De\hat z_j,\De\hat z_k]=i\h J_{jk},
\quad j,k=\overline{1,2n}, \]
which follow from (\ref{1.kom}) and (\ref{2.kom}), we find
\[ -i\h\lan\dot b(t),\De\hat z\ran+i\h\lan\De\hat z,\FH_{zz}(t)Jb(t)\ran=0 .\]
Hence, we have
\begin{equation}
\dot b=\FH_{zz}(t,\Psi_0)Jb.\label{kak80}
\end{equation}
Denote $b(t)=-Ja(t)$.  Then we obtain for the determination of the $2n$-space
vector $a(t)$ from (\ref{kak80})
\begin{equation}
\dot a=J\FH_{zz}(t,\Psi_0)a,\label{kak9}.
\end{equation}
We call the set of equations (\ref{kak9}), by analogy with the linear case
\cite{Mas4}, a {\em set of equations in variations}.

Thus, the operator
\begin{equation}
\hat a(t)=\hat a(t,\Psi_0)=N_a\lan b(t),\De\hat z\ran=
N_a\lan a(t),J\De\hat z\ran\label{kak6a}
\end{equation}
is a symmetry operator for equation (\ref{kak5}) if the vector
$a(t)=a(t,\Psi_0)$ is a solution of the equations in variations
(\ref{kak9}).

For each given solution $Z(t,\h)$ of the Hamilton--Ehrenfest
equations (\ref{bbst2.4}), we can find $2n$ linearly independent solutions
$a_k(t)\in \Bbb C^{2n}$ to the equations in variations (\ref{kak9}). Since
to each $2n$-space vector $a_k(t)$ corresponds an operator
$\hat a_k(t,\Psi_0)$, we obtain $2n$ operators $n$ of which commutate
with one another and form a complete set of symmetry operators for
equation (\ref{kak5}).

Now we turn to constructing the basis of solutions to equation
(\ref{kak5}) with the help of the operators $\hat a_k(t,\Psi_0)$. Equation
(\ref{kak5}) is a (linear) Schr\"odinger equation with a quadric
Hamiltonian and admits solutions in the form of Gaussian wave packets
\begin{equation}
\Phi(\vec x,t,\Psi_0)=N_\h\exp\Bigl\{\frac
i\h\Bigl[S(t,\h)+ i\phi_0(t)+i\h\phi_1(t)+ \lan\v P(t,\h),\De\vec
x\ran+\frac 12\lan\De\vec x,Q(t)\De\vec x\ran
\Bigr]\Bigr\},\label{kak14}
\end{equation}
where the real phase $S(t,\h)$ is defined in (\ref{kak15}), $N_\h$ is a
normalized constant, and the real functions $\phi_0(t)$ and $\phi_1(t)$ and
the complex $n\times n$ matrix $Q(t)$ are to be determined.

\begin{rem} Asymptotical solutions in the form of Gaussian packets
{\rm(\ref{kak14})} for equations with an integral nonlinearity of
more general form than {\rm(\ref{bbst1.1})} were constructed in
{\rm\cite{stud}}. In this case, the Hamilton--Ehrenfest equations
depend substantially on the initial condition for the starting
nonlinear equation.
\end{rem}

Substitution of (\ref{kak14}) into (\ref{kak5}) yields
\begin{eqnarray*}
& \Phi\Bigl\{\dot S(t,\h)+i\dot\phi_0(t)+i\h\dot\phi_1(t)+
\lan\dot{\vec P}(t,\h),\De\vec x\ran-
\lan\vec P(t,\h),\dot{\vec X}(t,\h)\ran+
\dac 12\lan\De\vec x,\dot Q(t)\De\vec x\ran-\\
& -\lan\De\vec x,Q(t)\dot{\vec X}(t,\h)\ran-\dot S(t,\h)+
\lan \v P(t,\h), \dot {\v X}(t,\h)\ran+
\lan\dot {\v X}(t,\h), Q(t)\De\v x\ran-
\lan \dot {\v P}(t,\h),\De\vec x\ran+\\
&+\dac 12\big\{\lan\De\vec x, \FH_{xx}(t,\Psi_0)\De\vec x\ran+
\lan\De\vec x,\FH_{px}(t,\Psi_0)Q(t)\De\vec x\ran+\\
& +\lan[-i\h\nabla+Q(t)\De\vec x],\FH_{px}(t,\Psi_0)\De\vec x\ran
+\lan[-i\h\nabla+Q(t)\De\vec x],\FH_{pp}(t,\Psi_0)
[-i\h\nabla+Q(t)\De\vec x]\ran\big\}\Bigr\}=0.
\end{eqnarray*}
Equating the coefficients at the terms with the same powers of the parameter
$\h$ and the operator $\De\vec x$, we obtain
\begin{eqnarray*}
(\De\vec x)^0\h^0: && i\dot\phi_0(t)=0;\\
(\De\vec x)^0\h^1: && i\dot\phi_1(t)+\dac{-i}2\Sp[\FH_{px}(t,\Psi_0)+
\FH_{pp}(t,\Psi_0)Q(t)]=0;\\
(\De\vec x)^1\h^0: && \lan\De\vec x,0\ran=0;\\
(\De\vec x)^2\h^0: && \lan\De\vec x,[\dot Q(t)+\FH_{xx}(t,\Psi_0)+
\FH_{xp}(t,\Psi_0)Q(t)+Q(t)\FH_{px}(t,\Psi_0)+\\
&& \quad +Q(t)\FH_{pp}(t,\Psi_0)Q(t)]\De\vec x\ran=0.
\end{eqnarray*}
As a result we have
\begin{eqnarray}
&& \phi_0(t)=0,\label{kak15a}\\
&& \phi_1(t)=\dac 12\dil_0^t \Sp[\FH_{px}(t)+\FH_{pp}(t)Q(t)]dt.
\label{kak16}
\end{eqnarray}

The matrix $Q(t)$ is determined from the Riccati type equation
\begin{equation}
\dot Q(t)+\FH_{xx}(t)+Q(t)\FH_{px}(t)+\FH_{xp}(t)Q(t)+
Q(t)\FH_{pp}(t)Q(t)=0.\label{kak17}
\end{equation}
Thus, the construction of a solution to equation (\ref{kak5}) in the form
of the Gaussian packet (\ref{kak14}) is reduced to solving the set
of ordinary differential equations (\ref{kak17}).

Let us now construct the Fock basis of solutions to the (linear)
Hartree equation in the trajectory-coherent approximation (\ref{kak5}).
This is the first step in constructing the solution to recurrent
equations (\ref{kak5})--(\ref{bbst3.10}).

Consider the properties of the symmetry operators $\hat a_k(t)$ (\ref{kak6a})
of the zero-order associated Schr\"odinger equation (\ref{kak5}),
which are necessary to construct the Fock basis.

\begin{demo} \rm Let $a_1(t)$ and $a_2(t)$ be two solutions of the
equations in variations and $\hat a_1(t)$ and $\hat a_2(t)$ be the
respective symmetry operators of equation (\ref{kak5}), defined in
(\ref{kak6a}). Then the equality
\begin{equation}
[\hat a_1(t),\hat a_2(t)]=i\h N_1N_2\{a_1(t),a_2(t)\}=
i\h N_1N_2\{a_1(0),a_2(0)\}\label{kak11}
\end{equation}
is valid.
\end{demo}
Actually, upon direct checking we are convinced that
\begin{eqnarray*}
& [\hat a_1(t),\hat a_2(t)]=N_1N_2[\lan a_1(t),J\De\hat z\ran,
\lan a_2(t),J\De\hat z\ran]=\\
& =i\h N_1N_2\lan Ja_1(t),JJa_2(t)\ran=\\
& =i\h N_1N_2\lan a_1(t),JJJa_2(t)\ran=i\h N_1N_2\lan a_1(t),
Ja_2(t)\ran=\\
& =i\h N_1N_2\{a_1(t),a_2(t)\}.
\end{eqnarray*}
Here, we have used the rules of commutation for the operators $\De\hat z$.
The skew scalar product holds and, hence, the statement is proved.

\begin{rem}  If the initial conditions for the equations in
variations are taken such that
\begin{eqnarray}
& {}\{a_j(0),a_k(0)\}=\{a_j^*(0),a_k^*(0)\}=0, \qquad
\{a_j(0),a_k^*(0)\}=id_k\de_{kj},\label{kak12} \\
& d_k>0, \qquad k,j=\overline{1,n},\nonumber
\end{eqnarray}
and $N_k=1/\sqrt{\h d_k}$, then the following canonical commutation
relations for the boson operators of {\rm``creation''} $(\hat
a_k^+(t))$ and {\rm``annihilation''} $(\hat a_k(t))$ are valid:
\begin{equation} [\hat a_k(t),\hat a_j(t)]=[\hat a_k^+(t),\hat
a_j^+(t)], \qquad [\hat a_k(t),\hat a_j^+(t)]=\de_{kj}.\label{kak13}
\end{equation}
The simplest example of initial data satisfying the conditions
{\rm(\ref{kak12})} is
\begin{equation}\begin{array}{l}
a_1(0)=(b_1,0,\dots,0,1,0,\dots);\\
a_2(0)=(0,b_2,\dots,0,0,1,\dots);\\
\dotfill\end{array}\label{primer}
\end{equation}
Here, we have $d_k=2\Im\,b_k>0$, $k=\overline{1,n}$.
\end{rem}

\begin{teo}  The function
\begin{eqnarray}
\lefteqn{ |0,t\ran=|0,t,\Psi_0\ran=\dac{N_\h}{\det C(t)}\times }\cr
&\times\exp\Big\{\dac i\h\Big[S(t,\h)+
\lan\vec P(t,\h),\De\vec x\ran+\frac 12\lan\De\vec x,Q(t)
\De\vec x\ran\Big]\Big\},\label{ros29}
\end{eqnarray}
where $N_\h=[(\pi\h)^{-n}\det D_0]^{1/4}$
is a {\rm``vacuum''} state for the operators $\hat a_j(t)$, such that
\begin{equation}
\hat a_j(t)|0,t\ran = 0, \quad j=\overline{1,n}.\label{a14}
\end{equation}
\end{teo}
{\bf Proof.}  Actually, substituting (\ref{kak6a}) and (\ref{ros29}) into
(\ref{a14}), we get
$$
|0,t\ran [\lan \vec Z_j(t), Q(t)\De \vec x \ran-
\lan \vec W_j(t), \De \vec x \ran ]=0,
$$
since we have
$$
Q(t) \vec Z_j(t) = B(t)C^{-1}(t) \vec Z_j(t) = \vec W_j(t).
$$

Recollect that from the fact that the matrix $D_0$ is positive
definite and diagonal follows $\det C(t)\ne0$, and so the matrix $\Im Q(t)$
is positive definite as well (see Appendix A).

Let us define the denumerable set of states $|\nu,t\ran$ as a result of
the action of the ``creation'' operators upon the ``vacuum'' state
$|0,t\ran$:
\begin{equation}
|\nu,t\ran=|\nu,t,\Psi_0\ran=\frac 1{\nu!}(\hat a^+(t,\Psi_0))^\nu
|0,t,\Psi_0\ran=\prod_{k=1}^n\frac 1{\nu_k!}(\hat a_k^+(t,\Psi_0))^{\nu_k}
|0,t,\Psi_0\ran.\label{deist}
\end{equation}

By analogy with the linear theory ($\vk=0$), we call the functions
$|\nu,t\ran$ (\ref{deist}) {\em quasi-classical trajectory-coherent states}
and consider their simplest properties.

\begin{demo} The relationships
\begin{eqnarray}
&& \hat a_k|\nu,t\ran =\sqrt{\nu_k}\,|\tilde\nu_k^{(-)},t\ran,\cr
&& \hat a_k^+|\nu,t\ran=\sqrt{\nu_k+1}\,|\tilde\nu_k^{(+)},t\ran,\\[2pt]
&& \tilde\nu_k^{(\pm)}=(\nu_1\pm\de_{1,k},\nu_2 \pm
\de_{2,k},\dots,\nu_n\pm\de_{n,k})\nonumber
\end{eqnarray}
are valid.
\end{demo}
Actually, we have
$$
[\hat a_j,(\hat a_k^+)^{\nu_k}] = \nu_k (\hat a_k^+)^{\nu_k-1}\de_{j,k}.
$$
It follows that
\begin{eqnarray*}
\hat a_j|\nu,t\ran&=&\displaystyle\prod_{k=1}^{n}\frac{1}{\sqrt{\nu_k!}}
[\hat{a}_j,(\hat a_k^+)^{\nu_k}]|0,t\ran=\prod_{k=1}^n\frac{\nu_j}
{\sqrt{\nu_k !}}(\hat{a}_k^+)^{\nu_k-\de_{k,j}}| 0,t \ran = {}\\
&=&\sqrt{\nu_j}\displaystyle\prod_{k=1}^n\frac{1}{\sqrt{(\nu_k- \de_{k,j})!}}
(\hat a_k^+)^{\nu_k-\de_{k,j}}|0,t\ran=\sqrt{\nu_j}|\tilde\nu_k^{(-)},t\ran;\\
\hat a_j^+|\nu,t\ran&=&\displaystyle\prod_{k=1}^n\frac{1}{\sqrt{\nu_k!}}
(\hat{a}_k^+)^{\nu_k+\de_{k,j}}|0,t\ran={}\\
&=&\displaystyle\prod_{k=1}^n\frac{\sqrt{\nu_j+1}}{\sqrt{(\nu_k+
\de_{k,j})!}}(\hat{a}_k^+)^{\nu_k+\de_{k,j}}|0,t\ran=
\sqrt{\nu_j+1}|\tilde\nu_k^{(+)},t\ran,
\end{eqnarray*}
and thus the statement is proved.

\begin{demo} The states  $|\nu,t,\Psi_0\ran$ with $t\in \BR$ and
$\Psi_0 \in \CP_\h^0$ form a set of orthonormal functions:
\begin{equation}
\lan \Psi_0,t,\nu'|\nu,t,\Psi_0\ran = \de_{\nu,\nu'},\quad \nu,\nu' \in
\BZ_+^n.
\end{equation}
\end{demo}
Let us consider the expression
$$
\lan \Psi_0,t,\nu'|\nu ,t,\Psi_0\ran = \frac{1}{\sqrt{\nu'!\nu !}}
\lan \Psi_0,t,0| \hat{\vec a}{\,}^{\nu'}(t,\Psi_0)[\hat{\vec a}{\,}^+]^\nu
(t,\Psi_0)|0,t,\Psi_0\ran.
$$
Commuting the operators of ``creation'' and ``annihilation'' in view of
commutation relations (\ref{kak13}) and using relationship
(\ref{a14}), we obtain
\begin{equation}
\lan t,\nu'|\nu,t\ran =\lan t,0|0,t\ran\de_{\nu,\nu'}.
\end{equation}
Then we calculate
\begin{equation}
\lan t,0|0,t\ran = \frac{N_\h^2}{|\det C(t)|}
\int\exp[-\frac{2}{\h}\Im S(\vec{x},t)]d\vec{x}.\label{kaka1}
\end{equation}
In view of (\ref{ros24}) and the explicit form of the complex phase in
(\ref{ros29}), we have
$$
\Im S(\vec{x},t) = \frac{1}{2}\lan\De\vec{x},\Im Q(t)\De\vec{x} \ran.
$$
The matrix $\Im Q(t)$ is real and positive definite; hence, the matrix
$\sqrt{\Im Q(t)}$ does exist, such that
$$
\det \sqrt{\Im Q(t)} = \frac{\sqrt{\det D_0}}{|\det C(t)|}.
$$
Let us perform in the integral of (\ref{kaka1}) the change
$$
\vec\xi = \frac{1}{\sqrt{\h}} \sqrt{\Im Q(t)} \De \vec{x}
$$
and then obtain
$$
\lan t,0|0,t\ran = \frac{N_\h^2}{\sqrt{\det D_0}}
\h^{n/2} \int e^{-\vec\xi^2} d \vec\xi
= \frac{(\pi\h)^{n/2}N_\h^2}{\sqrt{\det D_0}}=1
$$
since $\det D_0=\prod\limits^{n}_{k=1}\Im b_k$.  Thus, the functions
$|\nu,t,\Psi_0\ran$ (\ref{deist}) form the Fock basis of solutions to
equation (\ref{kak5}), Q.E.D.

\begin{teo} Let the symbols of the operators $\hat\CH(t)$ and $\hat V(t,\Psi)$
satisfy the conditions of Supposition \ref{assum1}. Then for any
$\nu\in\BZ^n_+$ the function
\begin{equation}
\Psi_\nu(\v x,t,\h)=|\nu,t\ran,\label{evol5}
\end{equation}
where the functions $|\nu,t\ran$ are defined by formula
{\rm(\ref{deist})}, is an asymptotical $($accurate to $O(\h^{3/2})$,
$\h\to0)$ solution to the Hartree type equation {\rm(\ref{bbst1.1})}
with the initial conditions
\begin{equation}
\Psi_\nu(\v x,t,\h)\big|_{t=0}=|\nu,t\ran|_{t=0}. \label{evoln}
\end{equation}
\end{teo}

\section{The principal term of the quasi-classical asymptotic of the
Hartree type equation}

The solution of the Cauchy problem (\ref{bbst1.1}), (\ref{evoln}) is a
special case of the quasi-classically concentrated solutions of equation
(\ref{bbst1.1}). However, in the case of arbitrary initial conditions
(\ref{bbst1.7}) belonging to the class $\CP_\h^t$, the functions $|\nu,t\ran$
are not asymptotical solutions of the Hartree type equation
(\ref{bbst1.1}).  This is a fundamental difference between the complex
germ method for the Hartree type equation (\ref{bbst1.1}), being
developed here, and a similar method developed for linear
equations \cite{Mas4, Mas4e, Bagre}.  The coefficients of the
Hartree type equation in the trajectory-coherent approximation
(\ref{kak5}) depend on the initial condition (\ref{bbst1.7}) since they
are determined by the solutions of the set of Hamilton--Ehrenfest
equations. It follows that among the whole set of solutions to equation
(\ref{kak5}) only one (satisfying the condition
$\Psi(\v x,t,\h)|_{t=0}=\Psi_0(\v x,\h)$) will be an asymptotical
(accurate to $O(\h^{3/2})$) solution to equation (\ref{bbst1.1}).
However, the Fock basis (TCS's) $|\nu,t\ran$ makes it possible to
construct an asymptotical solution to equation
(\ref{bbst1.1}) with any degree of accuracy in $\h^{1/2}$, $\h \to 0$
in an explicit form which would satisfy the initial condition
(\ref{bbst1.7}).

Let us illustrate in more detail the relation of the solutions
of the associated linear Schr\"odinger equation to the solution of the
Hartree type equation.  To do this, we construct the Green function of
the Cauchy problem for the zero-order associated Schrodinger equation.
Although the Green function $G^{(0)}(\v x, \v y,t,s)$ for quadric quantum
systems is well known \cite{Manko, 75, 76, 77}, we give for completeness
its explicit form, as convenient to us.  This function will allow us to
demonstrate explicitly the nontrivial dependence of the evolution operator
of the associated linear equation on the initial conditions for the starting
Hartree type equation.

By definition we have
\begin{equation}
\begin{array}{c}
[-i\h\pa_t + \hat\FH_0 (t,\Psi_0)]
G^{(0)}(\vec x,\vec y,t,s,\Psi_0)=0,\\[4pt]
\lim\limits_{t\to s} G^{(0)}(\vec x,\vec y,t,s,\Psi_0)=\de (\vec x-\vec y),
\end{array}\label{A.1}
\end{equation}
where the operator $\hat\FH_0$ is defined in (\ref{bbst3.5}). We make
use of the simplifying assumption that
\begin{equation}
\det\FH_{pp} (s) \neq 0, \quad \det \Big\|\frac {\pa p_k(t,z_0)}
{\pa p_{0j}}\Big\| \neq 0.\label{A.2}
\end{equation}

If the condition (\ref{A.2}) is not valid, the solution of the problem
can be found following the work \cite{76, 77}.  For the problem under
consideration, exact solutions of the Schr\"odinger equation
(\ref{kak5}) are known: these are the functions $|\nu,t,\Psi_0\ran$
(\ref{deist}) that form a complete set of functions.  Thus we have
\begin{equation}
G^{(0)}(\vec x,\vec y,t,s,\Psi_0)=\sum_{|\nu|=0}^\infty
\Phi_\nu(\vec x,t,\h)\Phi^*_\nu(\vec y,s,\h),\label{A.3}
\end{equation}
where
\[ \Phi_\nu(\vec x,t,\h)=|\nu,t,\Psi_0\ran. \]
Details of similar calculations can be found, for instance, in \cite{Manko}.
However, for our purposes the following approach seems to be convenient.

Let us carry out an $\h^{-1}$ Fourier transform in equation (\ref{A.1}).
For the Fourier transform of the Green function
\begin{equation}
\tilde G^{(0)}(\v p,\v y,t,s,\Psi_0)=\int_{{\Bbb R}^n}
\frac{d\v x}{(2\pi i\h)^{n/2}} G^{(0)}(\v x,\v y,t,s,\Psi_0)
\exp\Big\{-\frac i\h \lan\v x,\v p\ran\Big\}\label{A.4}
\end{equation}
we obtain
\begin{equation}\begin{array}{l}
[-i\h\pa_t +\tilde{\FH}_0 (\hat{\v p},\hat{\v x},t,\Psi_0)]
\tilde G^{(0)}(\vec p,\vec y,t,s,\Psi_0)=0,\\[6pt]
\lim\limits_{t\to s} \tilde G(\v p,\v y,t,s,\Psi_0)=
\dac 1{(2\pi i\h)^{n/2}}\exp \Big\{-\dac i\h
\lan \vec p,\vec y\ran \Big \}.\end{array}\label{A.5}
\end{equation}
Here, $\hat{\vec p}=\vec p$ and $\hat {\vec x} = i\h \frac{\pa}{\pa\vec
p}$ and the symbols of the operators of equations (\ref{A.1}) and
(\ref{A.5}) coincide:
\[ \tilde\FH_0 (\vec p,\vec x,t) = \FH_0 (\vec p,\vec x,t).\]

Equation (\ref{A.5}) coincides to notations with (\ref{kak5}) and,
hence, admits solutions of type (\ref{kak14})
\begin{equation}
\tilde G^{(0)}(\vec p,\vec y,t,s,\Psi_0)= \exp \Big\{-\frac i\h\Big[
S_0(t,s,\v y) + \lan\vec G(t,s,\v y),\De\vec p\ran +\frac 12\lan\De \vec p,
\tilde Q(t,s,\v y)\De\vec p\ran\Big]\Big\},\label{A.6}
\end{equation}
where $\De \vec p=\vec p-{\v P}(t,\h)$. Here, the functions
$S_0(t,s,\v y)=S_0(t)$, $\v G(t,s,\v y)=\v G(t)$ and
$\tilde Q(t,s,\v y)=\tilde Q(t)$ are to be determined and, according to
(\ref{A.5}), satisfy the initial conditions
\begin{equation}
\lim_{t\to s}\tilde Q (t,s,\v y) =0, \quad \lim_{t\to s}\vec G(t,s,\v y) =
\vec y , \quad \lim_{t\to s} S_0(t,s,\v y) =\lan \vec p_0,\vec y\ran.
\label{A.7}
\end{equation}
Substituting (\ref{A.6}) into (\ref{A.5}), we write
\begin{eqnarray*}
\lefteqn{\tilde G^{(0)}(\vec p,\vec y,t,s,\Psi_0)\Big\{-\dot S_0(t)-
\lan\dot{\vec G}(t),
\De\vec p(t)\ran+\lan\vec G(t),\dot{\vec P}(t,\h)\ran-\frac 1 2
\lan{\De\vec p}, \dot{\tilde Q(t)}\De \vec p\ran +}\\
&\hspace{-12pt}+ \lan \dot{\vec P}(t,\h), \tilde Q (t) \De \vec p
\ran -\dot S(t,\h)-\lan \dot{\v P}(t,\h),(\vec G(t) + \tilde Q(t)
\De \vec p-\vec P(t,\h))\ran +\lan \dot{\v X}(t,\h), \De \vec p \ran +\\
&+\dac 1 2 \big[\lan(\vec G(t)+\tilde Q(t)\De\vec p-{\v P}(t,\h)),
\FH_{xx}(t)(\vec G(t) + \tilde Q(t) \De \vec p-{\v P}(t,\h) ) \ran+\\
&+\lan(\tilde G(t)+\tilde Q(t)\De\vec p-{\v P}(t,\h)),\FH_{xp}(t)\De\vec
p\ran+\lan\De\vec p,\FH_{pp}(t)\De \vec p \ran+\\
&\hspace{-12pt}+\lan\De\vec p,\FH_{px}(t)\big(\tilde G(t)+\tilde Q(t)
\De\vec p-{\v P}(t,\h)\big)\ran\big]+\dac {i\h}2 \Sp\,[\FH_{xx}(t)
\tilde Q(t)+\FH_{xp}(t)]\Big\}=0.
\end{eqnarray*}
Equating the terms with the same powers of $\De\vec p$, we obtain the
following set of equations:
\begin{eqnarray}
&&\hspace{-24pt}-\dot{\tilde Q}=\FH_{px}(t)\tilde Q+\tilde Q\FH_{xp}(t)
+\tilde Q\FH_{xx}(t) \tilde Q+ \FH_{pp} (t) =0, \label{A.8}\\
&&\hspace{-24pt}-(\dot{\vec G}-\dot{\v P}(t,\h)) +\tilde Q(t) \FH_{xx}(t)
(\vec G-{\v P}(t,\h)) +\FH_{px} (t) (\vec G -{\v P}(t,\h))= 0, \label{A.9}\\
&&\hspace{-24pt}-\dot S_0 + \lan {\v X}(t,\h),\dot{\v P}(t,\h)\ran -
\dot S(t,\h) + \frac {i\h}2 \Sp~[\FH_{xx}(t)\tilde Q(t) +\FH _{xp}(t)]+\cr
&&+\frac 12\lan(\vec G(t)-{\v P}(t,\h)),\FH_{xx}(t)(\vec G(t)-
{\v P}(t,\h))\ran =0 \label{A.10}
\end{eqnarray}
with the initial conditions (\ref{A.7}).

Let $\tilde B(t)$ and $\tilde C(t)$ be solutions of equations in
variations (\ref{kak19}) with the initial conditions
\begin{equation}
\tilde B(t)|_{t=s}=B_0(s),\quad \tilde C(t)|_{t=s} =0,\quad
B_0^t(s)=B_0(s)\label{A.11}
\end{equation}
and the matrix $\Im\,B_0(s)$ be positive definite.

In view of (\ref{A.2}), the solution to the Cauchy problem
(\ref{kak19}), (\ref{A.11}) will then have the form
\begin{equation}
\tilde B(t)=\la^t_4(\De t)B_0(s), \quad \tilde C(t) =-\la^t_3(\De t)B_0(s),
\quad \De t=t-s,\label{A.12}
\end{equation}
where the matrices $\la^t_3 (t)$ and $\la^t_4 (t)$ are defined in (\ref{A.13}).
The matrix
\begin{equation}
\tilde Q(t) =\tilde C(t) \tilde B^{-1}(t) =
-\la^t_3(\De t)(\la_4^{-1}(\De t))^t\label{A.14}
\end{equation}
will then satisfy equation (\ref{A.8}) with the initial conditions
(\ref{A.7}).

Provided that (\ref{A.2}) is valid, from (\ref{B.22}) and (\ref{B.20})
follows
\begin{equation}
\vec G (t) =\big( \tilde B^{-1}(t) \big)^t \,B_0^t(s)\,(\vec y -\vec x_0)+
{\v X}(t,\h) = \la _4^{-1}(\De t)(\vec y -\vec x_0)+{\v X}(t,\h).
\end{equation}
In a similar manner, we obtain for $S_0$
\begin{eqnarray}
S_0(t,s,\h)&\hspace{-12pt}=&\hspace{-12pt} S(t,\h) -S(s,\h)
+\frac {i\h}2\int\limits_s^t \,d\tau \Sp~[\FH_{xp}(\tau)+\FH_{xx}(\tau)
\tilde Q(\tau)]+\cr
&+&\dac 1 2\int\limits_s^t d\tau\lan(\vec G(\tau)-{\v
X}(\tau,\h)),\FH_{xx}(\tau) (\vec G(\tau)-{\v X}(\tau,\h))\ran+\lan \vec
p_0,\vec y\ran.\label{A.15}
\end{eqnarray}

In view of (\ref{B.22}) and Liouville's lemma \ref{liuvil}, we obtain
\begin{eqnarray}
&\dac 1 2\int\limits_s^t \,d\tau\, \Sp~[\FH_{xp}(\tau)+\FH_{xx}
(\tau) \tilde Q(\tau) ]=\frac 1 2 \ln \det \tilde B^{-1} (\tau) |_s^t=\cr
&=\dac 1 2 \ln\Big(\frac{\det B_0(s)}{\det\tilde B(t)}\Big) =-~
\frac 1 2 \ln\det\la _4(\De t).\label{A.16}
\end{eqnarray}
To calculate the last integral in (\ref{A.16}), we use relationship
(\ref{B.28}) and, in view of (\ref{A.15}), we get
\begin{eqnarray}
&\dac 1 2 \int\limits_s^t d\tau\lan (\vec G(\tau)-{\v X}(\tau,\h)),
\FH_{xx}(\tau) (\vec G(\tau)-{\v X}(\tau,\h))\ran=\cr
&=\dac 1 2\lan(\vec y-\vec x_0),\la_2(\De t)\la^{-1}_4(\De t)
(\vec y-\vec x_0)\ran,\label{A.17}
\end{eqnarray}
where the matrix $\la^t_2 (t)$ is defined in (\ref{A.13}).  Hence, we
have
\begin{eqnarray}
S_0(t,s,\h)&=&S(t,\h) -S(s,\h)-\dac {i\h}2 \ln(\det \la_4(\De t))+\cr
&+&\dac 12 \lan(\vec y-\vec x_0),\la_2(\De t)\la^{-1}_4(\De t)(\vec y-
\vec x_0)\ran+\lan\vec p_0,\vec y\ran.\label{A.18}
\end{eqnarray}
Substituting (\ref{A.19}), (\ref{A.15}), and (\ref{A.14}) into (\ref{A.6}),
we obtain the well-known expression (see, e.g., \cite{76})
\begin{eqnarray}
\lefteqn{\hspace{-6pt}\tilde G^{(0)}(\vec p,\vec y,t,s,\Psi_0)\!=\!\dac{1}
{(2\pi i\h)^{n/2}}\frac 1{\sqrt{\det\la_4(\De t)}}\exp\!\Big\{\!\!-\frac i \h
(S(t,\h) -S(s,\h)-}\cr
&-\dac i{2\h}\lan(\vec y-\vec x_0),\la_2(\De t)\la^{-1}_4(\De t)
(\vec y-\vec x_0)\ran-\frac i \h\lan \vec p_0,\vec y\ran -\frac i\h
\lan \De \vec p,{\v X}(t,\h)\ran-\cr
&-\dac i \h\lan\De\vec p,\la_4^{-1}(\De t)(\vec y-\vec x_0)\ran
+\frac i{2\h}\lan\De\vec p,\la_4^{-1}(\De t)\la_3(\De t)\De\vec p\ran\Big\}.
\label{A.19}
\end{eqnarray}
Now we substitute (\ref{A.19}) into (\ref{A.4}) and make use of the
relationship
\begin{equation}
\int_{{\Bbb R}^n} d\vec x \exp \Big [ -\frac 1 2 \lan \vec x, \Gamma
\vec x\ran +\lan \vec b,\vec x\ran \Big]=\sqrt{(2\pi)^n\det
\Gamma^{-1}} \exp \Big\{ \frac {\lan \vec b, \Gamma^{-1} \vec b\ran} 2
\Big \},\label{A.20}
\end{equation}
in which we put $\Gamma =-(i/\h)\la_4^{-1}(\De t)\la_3(\De t)$,
$\vec b =-(i/\h)[\v X(t,\h)-\vec x+\la_4^{-1}(\De t)(\vec y-\vec x_0)]$.
We then obtain
\begin{eqnarray}
\lefteqn{G^{(0)}(\vec x,\vec y,t,s,\Psi_0)=
\dac 1{\sqrt{\det(-i2\pi\h\la_3(\De t))}}
\exp\Big\{\frac i \h\Big[S(t,\h)-S(s,\h)+}\cr
&+\lan\vec P(t,\h),\De\vec x\ran-\lan\vec p_0,(\vec y-\vec x_0)\ran-
\dac 1 2 \lan(\vec y-\vec x_0),\la_1(\De t)\la^{-1}_3(\De t)
(\vec y-\vec x_0)\ran-\cr
&-\dac 1 2\lan\De\vec x,\la_3^{-1}(\De t)(\vec y-\vec x_0)\ran-
\frac 1 2\lan\De\vec x,\la_3^{-1}(\De t)\la_4(\De t)\De\vec x\ran\Big] \Big\}.
\label{A.22}
\end{eqnarray}
Here, we used the relationships
\[ \la_1^t(t) \la_4(t) - \la_3^t(t)\la_2(t) ={\Bbb I}_{n\times n},
\qquad \la_3(t)\la_4^t(t) -\la_4(t)\la_3^t(t) = 0, \]
that follow immediately from (\ref{B.6}), (\ref{B.15}), and from the
definition of matriciant (\ref{A.13}).

Let us consider the limit of expression (\ref{A.22}) for $\De t=t-s\to0$.
We obtain
\begin{eqnarray*}
& \la_1(\De t) ={\Bbb I}_{n\times n}+O(\De t),\quad
\la_3^t(\De t)=-\FH_{pp}(s) \De t+O((\De t)^2),\\
& \la_3^{-1}(\De t) =-~\dac 1 {\De t} \FH^{-1}_{pp}(s)
+O((\De t)^0), \quad \la _4(\De t) ={\Bbb I}_{n\times n}+O(\De t),\\
& \la_2(\De t) =O(\De t).
\end{eqnarray*}
It follows that for short times we have (see, e.g., \cite{64})
\begin{eqnarray}
\lefteqn{\lim_{\De t\to 0} G^{(0)}(\vec x,\vec y,t,s,\Psi_0)=\dac 1
{\sqrt{\det(-i2\pi\h\De t\FH_{pp}(s))}}\times}\cr
&\times\exp\Big\{\dac i{2\h\De t}\lan (\vec x -\vec y), \FH^{-1}_{pp}(s)
(\vec x -\vec y)\ran + O(\De t ^0)\Big \}.
\end{eqnarray}
Thus we have proved

\begin{teo} Let the symbols of the operators $\hat\CH(t)$ and
$\hat V(t,\Psi)$ satisfy the conditions of Supposition \ref{assum1}.
Then the function
\begin{equation}
\Psi^{(0)}(\v x,t,\h)=\hat U^{(0)}(t,0,\Psi_0)\Psi_0,\label{evol6}
\end{equation}
where $\hat U^{(0)}(t,0,\Psi_0)$ is the evolution operator of the zero-order
associated Schr\"odinger equation {\rm(\ref{kak5})} with the kernel
$G^{(0)}(\v x,\v y,t,0,\Psi_0)$, is an asymptotical $($accurate to
$O(\h^{3/2})$, $\h\to0)$ solution of the Hartree type equation
{\rm(\ref{bbst1.1})} and satisfies the initial condition
\[ \Psi^{(0)}(\v x,t,\h)\big|_{t=0}=\Psi_0. \]
\end{teo}

\begin{rem} The principal term of the quasi-classical asymptotic
$\Psi^{(0)}(\v x,t,\h)$ will not change $($accurate to $O(\h^{3/2})$,
$\h\to0)$ if the phase function $S^{(N)}(t,\h)$ in the operator
$\hat\FH_0(t,\Psi_0)$ is substituted by its value $S^{2}(t,\h)$ for
$N=2$ and we restrict ourselves to the first terms in $\h\to0$ in the
phase trajectory $Z^{(2)}(t,\h)$, and in the other expressions
$Z^{(N)}(t,\h)$ is changed by $Z^{0}(t,\h)$.
\end{rem}

\section{Quasi-classically concentrated solutions of the Hartree type
equation}

Now we construct asymptotical solutions to the Hartree type equation
(\ref{bbst1.1}) with an arbitrary accuracy in powers of $\sqrt\h$.
To do this, we find asymptotical solutions to the associated linear
Schr\"odinger equation (\ref{bbst3.2b}) with an arbitrary accuracy in
powers of $\sqrt\h$.  Let us present an arbitrary initial condition
$\Phi_0(\vec x,\h)\in\CP_\h^0$
\begin{equation}
\Phi_0(\v x,\h)=\sum_{k=0}^N\h^{k/2}\Phi^{(k)}_0 (\v x,\h),
\label{kaka2}
\end{equation}
where
\[\Phi^{(k)}_0 (\v x,\h)\in\CC_\h^t(z_0,S_0).\]
Then for the recurrent associated linear equations
(\ref{kak5})--(\ref{bbst3.10}) we arrive at a Cauchy problem with
initial data:
\[ \Phi^{(k)}\big|_{t=0}=\Phi^{(k)}_0(\v x,\h) \quad
k=\overline{0,N}. \]
The solution to these recurrent equations can readily be
constructed as its expansion over the complete set of orthonormalized
Fock functions $|\nu,t\ran$ (\ref{a14}).  As a result we obtain
\begin{eqnarray}
&\displaystyle \Phi^{(0)}(\v x,t,\h)=\sum_{|\nu|=0}^\iy
|\nu,t,\Psi_0\ran\lan \Psi_0,0,\nu| \Phi^{(0)}_0(\v x,\h)\ran,\label{kaka3}\\
&\displaystyle \Phi^{(1)}(\v x,t,\h)=\sum_{|\nu|=0}^\iy |\nu,t,\Psi_0\ran
\lan \Psi_0,0,\nu|\Phi^{(1)}_0(\v x,\h)\ran-\cr
&-\dac i\h\sum_{|\nu|=0}^\iy |\nu,t,\Psi_0\ran
\inl_0^t d\tau\,\lan\Psi_0,\tau,\nu|\hat\FH_1(t,\Psi_0)\Phi^{(0)}(\v x,
\tau,\h)\ran,\label{kaka4}\\
&\displaystyle \Phi^{(2)}(\v x,t,\h)=\sum_{|\nu|=0}^\iy |\nu,t,\Psi_0\ran
\lan \Psi_0,0,\nu|\Phi^{(2)}_0(\v x,\h)\ran-\cr
&-\dac i\h\sum_{|\nu|=0}^\iy |\nu,t,\Psi_0\ran
\inl_0^t d\tau\,\lan\Psi_0,\tau,\nu|\hat\FH_1(t,\Psi_0)\Phi^{(1)}(\v x,
\tau,\h)\ran-\cr
&-\dac i\h\sum_{|\nu|=0}^\iy |\nu,t,\Psi_0\ran
\inl_0^t d\tau\,\lan\Psi_0,\tau,\nu|\hat\FH_2(t,\Psi_0)\Phi^{(0)}(\v x,
\tau,\h)\ran,\label{kaka40}\\
& \dots\dots\dots\dots\dots\dots\dots\dots\nonumber
\end{eqnarray}

Denote by $\hat{\cal F}^{(N)}(t,\Psi_0)$ the operator defined by the
relationship
\begin{equation}
 \hat{\cal F}^{(N)}(t,\Psi_0)\Phi(\vec x,t)=\dil_0^t d\tau\,
\hat U_0(t,\tau,\Psi_0)\hat\FH^{(N)}(\tau,\Psi_0)\Phi(\v x,\tau),\label{evol2}
\end{equation}
where $\hat U_0(t,\tau)$ is the evolution operator of the associated
Schr\"odinger equation (\ref{kak5}) and the following notation has been made:
\begin{equation}
\hat\FH^{(N)}(t,\Psi_0)=\sum_{k=1}^N \h^{k/2}\hat\FH_k(t,\Psi_0).
\label{evol3}
\end{equation}
Thus we have proved the following statement:

\begin{teo} Let the symbols of the operators $\hat\CH(t)$ and $\hat V(t,\Psi)$
satisfy the conditions of Supposition \ref{assum1}. Then the function
\begin{equation}
\Psi^{(N)}(\vec x,t,\h)=\sum_{k=0}^N \frac 1{k!}\Big\{
\frac {-i}\h\hat{\cal F}^{(N)}(t,\Psi_0)\Big\}^k \hat
U_0(t,0,\Psi_0)\Psi_0(\vec x,\h).\label{evol4},
\end{equation}
where $N\ge2$, is an asymptotical, accurate to $O(\h^{(N+1)/2})$, solution
of equation {\rm(\ref{bbst1.1})} and satisfies the initial condition
{\rm(\ref{bbst1.11}).}
\end{teo}

\section{The Green function and the nonlinear superposition principle}

Let us show that in the class of trajectory-concentrated functions for
the Hartree type equation (\ref{bbst1.1}) we can construct, with any
given accuracy in $\h^{1/2}$, the kernel of the evolution operator or
the Green function of the Cauchy problem for equation (\ref{bbst1.1}).
The explicit form of the quasi-classical asymptotics $\Psi^{(N)}(\vec
x,t,\h)$ (\ref{evol4}) makes it possible to obtain an expression for
the Green function $G^{(N)}(\vec x,\vec y,t,s,\Psi_0)$ valid on finite
time intervals $t\in [0,T]$. Actually, according to (\ref{evol4}), for
any function $\va(\vec x,\h)\in\CP_\h^0$, the solution of the Cauchy
problem with the initial condition
\begin{equation}
\Phi(\vec x,t,\h)|_{t=0} = \va(\vec x,\h)\label{6.2}
\end{equation}
for the associated linear Schr\"odinger equation (\ref{bbst3.4}) has the
form
\begin{eqnarray*}
\lefteqn{\Phi^{(N)}(\vec x,t,\h)=}\\[0.5ex]
&=\hat R^{(N)}(t,\Psi_0)\!\dil_{\BR^n}\! d\vec y G^{(0)}(\vec x,\vec
y,t,0,\Psi_0)\va(\vec y,\h)  + O(\h^{(N+1)/2})=\\[0.5ex]
&=\dil_{\BR^n}d\vec yG^{(N)} (\vec x,\vec y,t,0,\Psi_0)\va(\vec y,\h)+
O(\h^{(N+1)/2}),
\end{eqnarray*}
where
\begin{equation}
\hspace{-12pt}\hat R^{(N)} (t,\Psi_0)=\sum_{k=0}^N \frac 1{k!}
\Big\{\frac {-i}\h\hat{\cal F}^{(N)}(t,\Psi_0)\Big\}^k,\label{6.3}
\end{equation}
and the function $G^{(0)}(\vec x,\vec y,t,s,\Psi_0)$ is defined
in (\ref{A.22}).

It follows that
$$
G^{(N)}(\vec x,\vec y,t,0,\Psi_0) =
\hat R^{(N)}(t,\Psi_0) G^{(0)}(\vec x,\vec y,t,0,\Psi_0).
$$
Since $\hat R^{(N)}(0,\Psi_0)=1$, we have for an arbitrary $s<t$
\begin{equation}
G^{(N)}(\vec x,\vec y,t,s,,\Psi_0)=
\hat R^{(N)}(t,\Psi_0) G^{(0)}(\vec x,\vec y,t,s,\Psi_0)
\big(\hat R^{(N)}(s,\Psi_0)\big)^+\label{6.4}
\end{equation}
being the Green function of the Cauchy problem (\ref{6.2}) with $s\ne0$.
Obviously, for the functions $G^{(N)}(\vec x,\vec y,t,s,\Psi_0)$ the
following composition rule is valid:
\begin{eqnarray}
\int d\vec u G^{(N)}(\vec x,\vec u,t,\tau,\Psi_0)G^{(N)}(\vec u,\vec y,
\tau,s,\Psi_0)&=&G^{(N)}(\vec x,\vec y,t,s,\Psi_0)+\cr
&+&O(\h^{(N+1)/2}).\label{6.6}
\end{eqnarray}

Denoting by $\hat U^{(N)}(t,0,\Psi_0)$ the approximate evolution operator
of the linear equation (\ref{bbst3.4})
\[ \hat U^{(N)}(t,0,\Psi_0)\va(\vec x,\h)=\int d\vec y\,G^{(N)}
(\vec x,\vec y,t,0,\Psi_0)\va(\vec y,\h), \]
we obtain it from (\ref{6.4}) in the form of the T-ordered Dyson
expansion
\begin{equation}
\hat U^{(N)}(t,0,\Psi_0)=\sum_{k=0}^N\Big( -\frac i \h \Big)^k
\inl_{\De_k^>} d^k\tau\hat\FH_1(\tau_1,t,\Psi_0) \cdots
\hat\FH_1(\tau_k,t,\Psi_0)\hat U_0(t,0,\Psi_0).\label{6.7}
\end{equation}
Here, we have used the following notations \cite{64}: the domain of
integration is an open hypertriangle
$$ \De_k^> \equiv \{\tau \in [0,t]^k |t > \tau_1 > \tau_2
> \cdots > \tau_N >s\}; $$
the operator $\hat\FH_1(\tau,t,\Psi_0)$ is a perturbation operator
in the representation of the interaction
\begin{equation}
\hat\FH_1(\tau,t)=\hat U_0(t,\tau,\Psi_0)\hat\FH^{(N)}(\tau,\Psi_0)\hat U_0^+
(\tau,t,\Psi_0),\label{6.8}
\end{equation}
where $\hat\FH^{(N)}(t,\Psi_0)$ has been defined in (\ref{bbst3.5a}), and
$\hat U_0(t,s,\Psi_0)$ is the evolution operator of the associated linear
Schr\"odinger equation with the kernel $G^{(0)}(\vec x,\vec y,t,s,\Psi_0)$
(\ref{A.22}).

In view of Statement \ref{d1}, the action of operator (\ref{6.7}) on the
function $\va=\Psi_0(\v x,\h)$ determines the asymptotical solution
of the Cauchy problem (\ref{bbst1.1})--(\ref{bbst1.11}) for the Hartree type
equation (\ref{bbst1.1})
\begin{equation}
\Psi^{(N)}(\v x,t,\h)=\hat U^{(N)}(t,0,\Psi_0)\Psi_0(\v x,\h),
\quad \Psi_0(\v x,\h)\in\CP^0_\h.\label{evo}
\end{equation}
It follows that operator (\ref{6.7}) is an approximate evolution
operator for the Hartree type equation (\ref{bbst1.1}) in the class of
trajectory-concentrated functions.

For the constructed asymptotical solutions, from expression (\ref{evo})
immediately follows \cite{stud2, BBTSh}

\begin{teo} [nonlinear superposition principle]
Let $\Psi_1(\v x,t,\h,y_1^{(N)}(t,\h))$ be an asymptotical, accurate to
$O(\h^{(N+1)/2})$, solution of the Cauchy problem for the Hartree type
equation {\rm(\ref{bbst1.1})} with the initial condition
$\Psi_{01}(\v x,\h)$ and the function $\Psi_2(\v x,t,\h,y_2^{(N)}(t,\h))$
is a solution of the same problem with the initial condition
$\Psi_{02}(\v x,\h)$.  Then the solution of the Cauchy problem with the
initial condition
\[ \Psi_{03}(\v x,\h)=c_1\Psi_{01}(\v x,\h)+c_2\Psi_{02}(\v x,\h),
\qquad c_1,c_2=\rm const, \]
has the form
\begin{eqnarray*}
& \Psi_3(\v x,t,\h,y_3^{(N)}(t,\h))=
\hat U^{(N)}(t,0,\Psi_{03})\Psi_{03}(\v x)=\\
&= c_1\hat U^{(N)}(t,0,\Psi_{03})\Psi_{01}(\v x) +
c_2\hat U^{(N)}(t,0,\Psi_{03})\Psi_{02}(\v x)=\\
&=c_1\Psi_1(\v x,t,\h,
y_3^{(N)}(t,\h))+c_2\Psi_2(\v x,t,\h,y_3^{(N)}(t,\h)).
\end{eqnarray*}
Here, $y_k^{(N)}(t,\h)$ denotes the solution of the
Hamilton--Ehrenfest equations of order $N$,
$N\ge2$ {\rm(\ref{bbst2.4})} with an initial condition which is
determined from the functions $\Psi_{0k}(\v x,\h)$, $k=\overline{1,3}$,
respectively.
\end{teo}

\section{The one-dimensional Hartree type equation with a Gaussian potential}

Let us illustrate the above scheme for constructing asymptotical
solutions by the example of a nonlinear interaction with a Gaussian
potential \cite{stud3}.  By this example we shall show in an explicit
form how the procedure of constructing quasi-classically concentrated
solutions to the Hartree type equation necessarily leads to
Hamilton--Ehrenfest equations. Moreover, it becomes possible to elucidate
the ``nonlinearity'' of the generalized superposition principle for the
Hartree type equation.

We write equation (\ref{bbst1.1}) with a Gaussian potential for the
one-dimensional case as
\begin{equation} \label{gauss1}
\left\{ -i\h\pa_t +\frac{(\hat p)^2}{2m}+\vk V_0\dil_{-\iy}^{+\iy}dy\,
{\exp\Bigl[\frac{(x-y)^2}{2\gamma^2}\Bigr]
|\Psi(y,t)|^2}\right\}\Psi =0.
\end{equation}
In this case, for the class of functions $\CP_\h^t(S(t,\h),Z(t))$ in which
we shall find solutions to equation (\ref{gauss1}), in accordance with
(\ref{bbst1.5}), we find
\begin{eqnarray}
\lefteqn{\CP_\h^t\big(S(t,\h),Z(t)\big) =}\cr
&&\displaystyle=\biggl\{\Psi :\Psi(x,t,\h)=
   \va\Bigl(\frac{\De x}{\sqrt{\h}},t,\h\Bigr)
   \exp\Bigl[{\frac{i}{\h}(S(t)+P(t)\De x)}\Bigr]\biggr\}.\label{bbst1.2g}
\end{eqnarray}
Here, the function $\va(\xi ,t,\h )\in{\Bbb S}$ (Schwartz space) with
respect to the variable $\xi=\dac{\De x}{\sqrt{\h}}$ and depends
regularly on $\h$ with $\De x=x-x(t)$. The functions $S(t,\h)$ and
$Z(t)=(P(t),X(t))$ are real, depend regularly on $\h$, and are to be
determined.

Let us expand the exponential in equation (\ref{gauss1}) in a Taylor
series for $\De x=x-X(t)$, $\De y=y-X(t)$ and restrict ourselves to the
terms of the order of not above four in $\De x$ and $\De y$. In view of the
estimates (\ref{bbst1.10}), equation (\ref{gauss1}) will then
take the form
\begin{eqnarray}
&\biggl\{ -i\h\pa_t +\dac{P^2(t)}{2m}+\frac{P(t)\De\hat p}{m}+
\frac{\De\hat p^2}{2m}+\nonumber\\
&+\tvk V_0\Big[1- \dac{1}{2\gamma^2}(\De x^2-2\De
x\al_\Psi^{(1)}(t,\h)+\al_\Psi^{(2)}(t,\h))+\nonumber
\\&+\dac{1}{8\gamma^4}(\De x^4-4\De x^3\al_\Psi^{(1)}(t,\h)+6\De
x^2\al_\Psi^{(2)}(t,\h)- \nonumber\\
&-4\De x\al_\Psi^{(3)}(t,\h)+\al_\Psi^{(4)}(t,\h))\Big]\biggr\}\Psi=
O(\h^{5/2}),\label{gauss4}
\end{eqnarray}
where $\De{\hat p}=\hat p-P(t)$, $\tvk=\vk \|\Psi\|^2$, and
\begin{eqnarray*}
\al_\Psi^{(k)}(t,\h)=\frac{1}{\|\Psi\|^2}\dil_{-\iy}^{+\iy}
(\De y)^{k}{|\Psi(y,t)|^2}dy, \quad k=\overline{0,\iy}
\end{eqnarray*}
are the $k$-order moments centered about $X(t)$. Equation
(\ref{gauss4}) can be simplified if we make the change
\begin{eqnarray}
\Psi(x,t,\h)=\exp{\biggl\{-\frac{i}{\h}\dil_{0}^{t}
\Big[\frac{P^2(t)}{2m}-\tvk V_0+\frac{\tvk}{2\gamma^2}
V_0\sigma_{xx}(t,\h)\Big]dt\biggr\}} \Phi(x,t,\h), \label{gauss5}
\end{eqnarray}
where
$$
\sigma_{xx}(t,\h)=\al_\Psi^{(2)}(t,\h)=\frac{1}{\|\Psi\|^2}
\dil_{-\iy}^{+\iy}\De y^{2}{|\Psi(y,t)|^2}dy
$$
is the variance.  The function $\Phi(x,t,\h)$ belongs to the class
$\CP_\h^t\big(\tilde S(t,\h),Z(t)\big)$, where
$$
\tilde S(t,\h)=S(t,\h)-\dil_{0}^{t}
\Big[\frac{P^2(t)}{2m}+\tvk V_0 -\frac{\tvk}{2\gamma^2}
V_0\sigma_{xx}(t,\h)\Big]dt,
$$
and satisfies the equation
\begin{eqnarray}
&\biggl\{ -i\h\pa_t +\dac{P(t)\De\hat p}{m}+
\frac{\De\hat p^2}{2m}+
\tvk V_0\Big[- \frac{1}{2\gamma^2}(\De x^2-2\De
x\al_\Phi^{(1)}(t,\h))+\nonumber\\
&+\dac{1}{8\gamma^4}(\De x^4-4\De x^3\al_\Phi^{(1)}(t,\h)+6\De
x^2\al_\Phi^{(2)}(t,\h)- \nonumber\\
&-4\De x\al_\Phi^{(3)}(t,\h)+\al_\Phi^{(4)}(t,\h))\Big]\biggr\}\Phi=
O(\h^{5/2}).\label{gauss6}
\end{eqnarray}
Here, we have made use of
\[ \al_\Psi^{(k)}(t,\h)=\al_\Phi^{(k)}(t,\h), \qquad
k=\overline{1,N}. \]

We shall seek the approximate $(\bmod\h^{5/2})$ solution $\Phi(x,t,\h)$
to equation (\ref{gauss6}) in the form
$$
\Phi(x,t,\h)=\Phi^{(0)}(x,t)+\sqrt{\h}\Phi^{(1)}(x,t)+
\h\Phi^{(2)}(x,t)+\dots,
$$
where $\Phi^{(k)}(x,t)\in\CC_\h^t(S(t,\h),Z(t))$.
In equation (\ref{gauss6}) we equate the terms having the same estimate
in $\sqrt\h$ in the sense of (\ref{bbst1.10}). Denote by $\hat L_0$ the
operator
\begin{eqnarray*}
\hat L_0=-i\h\pa_t +\dac{1}{m}P(t)\De\hat p+
\frac{1}{2m}\De\hat p^2-\frac{\tvk V_0}{2\gamma^2}\De x^2;
\end{eqnarray*}
Earlier we have shown that $\hat L_0=\hat O(\h)$.  We then have
\begin{eqnarray}
\h^1 &:& (\hat L_0+\frac{\tvk}{2\gamma^2} \De x\al_{\Phi^{(0)}}^{(1)})
\Phi^{(0)}=0;  \label{gauss7}\\
\h^{3/2} &:& \sqrt{\h}(\hat L_0+\frac{\tvk}{2\gamma^2} \De
x\al_{\Phi^{(0)}}^{(1)})\Phi^{(1)}=-2\sqrt{\h}\dac{\tvk V_0}{\gamma^2}\De x
\Re \lan\Phi^{(0)}|\De x|\Phi^{(1)}\ran\Phi^{(0)},\cr
\h^2 &:& \h(\hat L_0+\frac{\tvk}{2\gamma^2} \De x\al_{\Phi^{(0)}}^{(1)})
\Phi^{(2)}=-2\h\dac{\tvk V_0}{\gamma^2}\De x \biggr\{\Big[2\Re\lan
\Phi^{(0)}|\De x|\Phi^{(2)}\ran+\cr
&{}&+\lan\Phi^{(1)}|\De x|\Phi^{(1)}\ran\Big]\Phi^{(0)}+
2\Re \lan\Phi^{(0)}|\De x|\Phi^{(1)}\ran\Phi^{(1)}\biggl\}-
\hat{\FH}(t)\Phi^{(0)},\nonumber
\end{eqnarray}
The function
\begin{eqnarray}
\lefteqn{\Phi^{(0)}(x,t)=\Phi_0^{(0)}(x,t)=\biggl(\frac{1}{|C(t)|}\biggr)^{1/2}\exp{
\biggl\{\frac{i}{\h}\Big(\frac{v^2t}{m}+v\De x+\frac12\De
x^2m\frac{\dot C(t)}{C(t)}\Big)\biggr\}} =}\cr
&\displaystyle=\biggl(\frac{1}{|C(t)|}\biggl)^{1/2}
\exp{\biggl\{\frac{i}{\h}\Big(\frac{v(x-x_0)}m
+\frac12\De x^2m\frac{\dot C(t)}{C(t)}\Big)\biggl\}}\nonumber
\end{eqnarray}
is a solution of equation (\ref{gauss7}). Here, we have used the fact
that $X(t)$ and $P(t)$ are solutions of the ordinary differential
equations
\begin{equation}
\left\{\begin{array}{l}
\dot{p}=0\\
\dot{x} =\dac{p}{m},\quad p(0)=p_0,x(0)=x_0,
\end{array}\right.\label{gauss8}
\end{equation}
and $C(t)$ denotes the complex function satisfying the equations
\begin{equation}
\left\{\begin{array}{l}
\dot B=\dac{\tvk V_0}{\gamma^2}C\\[12pt]
\dot C=\dac{B(t)}{m}.\end{array}\right.\label{gauss9}
\end{equation}
Equations (\ref{gauss8}) are Hamiltonian equations
with the Hamiltonian $\CH(p,x,t)=p^2/(2m)$ and their solution is
\[ P(t)=mv=p_0,\qquad X(t)=\frac{p_0}{m}t+x_0. \]
Similarly, equations (\ref{gauss8}) are Hamiltonian
equations for a harmonic oscillator with frequency
$$
\Omega=\sqrt{\frac{\tvk |V_0|}{m\gamma^2}},
$$
and their solution is
\begin{eqnarray*}
\left\{\begin{array}{l}
C(t) =c_1 \exp{\Big(-\sqrt{\dac{\tvk |V_0|}{m\gamma^2}}t\Big)}+c_2
\exp{\Big(+\sqrt{\dac{\tvk |V_0|}{m\gamma^2}}t\Big)}\\
\dot B(t)=m\dot C(t).\end{array}\right.
\end{eqnarray*}
For the initial conditions (\ref{primer})
\[ C(0)=1, \quad \dot B(0)=b,\quad \Im b>0 \]
two cases are possible:
\begin{equation}
\begin{array}{l}
1)~~\tvk V_0>0: \quad C(t)=\ch(\Omega t)+\dac b\Omega\sh(\Omega t),\\[12pt]
2)~~\tvk V_0<0:\quad C(t)=\cos(\Omega t)+\dac{b}{\Omega}\sin(\Omega t).
\end{array}
\label{gauss10}
\end{equation}
In the linear case ($\tvk=0$), the frequency $\Omega=0$ and equations
(\ref{gauss9}) become equations in variations for equations (\ref{gauss8}).
In view of (\ref{gauss9}), we find the variance of the coordinate $x$
in an explicit form:
\begin{eqnarray}
\lefteqn{\sigma_{xx}(t,\h)=}\cr
&&\displaystyle=\sqrt{\frac{m\Im b}{\pi \h}}\cdot\dil_{-\iy}^{+\iy}
\frac{\De {x^2}}{|C(t)|}\exp\Bigl[-\frac{m}{\h}\De
{x^2}\frac{\Im b}{|C(t)|^2}\Bigr]dx=\frac{|C(t)|^2
\h}{2m\cdot\Im b}. \label{gauss11}
\end{eqnarray}
Then we get
\begin{eqnarray*}
\lefteqn{\Psi_0^{(0)}(x,t,\h)=}\\
&&\displaystyle=\exp{\Big[-\frac{i}{\h}
\Big((\frac{v^2}{2m}+\tvk V_0)t+\tvk
V_0\frac{\h}{2m\cdot\Im b}\dil_{0}^{t}|C(t)|^2dt\Big)\Big]}
\Phi_0^{(0)}(x,t,\h).
\end{eqnarray*}
It can readily be noticed that
$\al_{\Psi^{(0)}}^{(2)}(t,\h)=\al_{\Phi^{(0)}}^{(2)}(t,\h)$.  Hence, from
(\ref{gauss10}) and (\ref{gauss11}) it can be inferred that for $\tvk
V_0<0$ the variance $\al_{\Phi^{(0)}}^{(2)}(t,\h)$ is limited in $t$,
i.e., we have $|\sigma_{xx}(t)|\le M$, $M=\const$, and for $\tvk V_0>0$
it increases exponentially.  In the limit of $\gamma\to 0$ and with
$V_0=(2\pi\gamma)^{-1/2}$, equation (\ref{gauss6}) becomes a nonlinear
Schr\"odinger equation, while in the case where $\tvk V_0<0$ ($\tvk V_0>0$)
it corresponds to the condition of existence (nonexistence) of solitons.
Note that if $\al_{\Phi^{(0)}}^{(1)}(t,\h)=0$, the equation for
the function $\Phi^{(0)}$ takes the form
\begin{equation}
\hat L_0\cdot\Phi^{(0)}=0\quad\label{gauss12}
\end{equation}
becoming a Schr\"odinger equation with a quadric Hamiltonian.  We
shall find the solution to equation (\ref{gauss12}) satisfying an
additional condition $\al_{\Phi^{(0)}}^{(1)}(t,\h)=0$. To do
this, we denote
\[ \hat a(t)=N_a(C(t)\De {\hat p}-B(t)\De x). \]
If $C(t)$ and $B(t)$ are solutions of equations
(\ref{gauss9}), the operator $\hat a(t)$ commutates with the operator
$\hat L_0$.  So the function
\[ \Phi_k^{(0)}=\frac{1}{k!}(\hat a^+(t))^k\Phi_0^{(0)} \]
will also be a solution of the Schr\"odinger equation (\ref{gauss6}).
Commuting the operators $\hat a^+(t)$ with the function
$\Phi_0^{(0)}(x,t,\h)$, we obtain the Fock basis of solutions for linear
equation (\ref{gauss12})
\begin{eqnarray*}
\lefteqn{\Phi_k^{(0)}(x,t)=\frac{1}{k!}N^k_a\Phi_0^{(0)}(x,t)(-i)^k[C^*(t)]^k
\Big[\h\frac\pa{\pa x}-\frac{2m\,\Im b}{|C(t)|^2}\De x\Big]^{k} 1=}\\
&=\dac{1}{k!}N^k_a\Phi_0^{(0)}(x,t)(-i)^k[C^*(t)]^k
\biggl(\frac{\sqrt{\h m\,\Im b}}{|C(t)|}\biggr)^k H_k
\biggl(\De x\frac{\sqrt{m\,\Im b}}{|C(t)|\sqrt\h}\biggr),
\end{eqnarray*}
where $H_n(\xi)$ are Hermite polinomials.  Determining $N_a$ from the
condition $[\hat a(t),\hat a^+(t)]=1$ and representing the solution of
the equations in variations as
\[ C(t)=|C(t)|\exp\big\{-i\Arg[C(t)]\big\}, \]
we get
\begin{equation}
\Phi_k^{(0)}(x,t)=\frac{1}{k!}(-i)^k\exp\big\{-ik\Arg[C(t)]\big\}
\biggl(\frac{1}{\sqrt{2}}\biggr)^k
H_k\biggl(\De x\frac{\sqrt{m\,\Im b}}{|C(t)|\sqrt{\h}}\biggr)^{k}
\Phi_0^{(0)}(x,t).\label{gauss13}
\end{equation}
Using the properties of Hermite polinomials, we can readily be
convinced that the mean $\al^{(1)}_{\Phi^{(0)}_k}(t,\h)=0$,
$k=\overline{0,\iy}$. Then we have
\begin{eqnarray*}
\lefteqn{\Psi_k^{(0)}(x,t,\h)=}\\
&&\displaystyle=\exp\Big\{-\frac{i}{\h}
\Big[\Big(\frac{v^2}{2m}+\tvk V_0\Big)t+\tvk
V_0\lan\Phi_k^{(0)}|\De x^2|\Phi_k^{(0)}\ran\Big]\Big\}
\Phi_k^{(0)}(x,t,\h).
\end{eqnarray*}
Similarly, we find
\[ \lan\Phi_k^{(0)}|\De x^2|\Phi_k^{(0)}\ran=\frac{1}{2^k k!\sqrt{\pi}}
\inl_{-\iy}^\iy\De x^2|\Phi_k^{(0)}(x,t)|^2dx=
\frac{\h|C(t)|^2(2k+1)}{2m\,\Im b} \]
and for functions $\Psi^{(0)}_k(t,\h)$ obtain
\begin{eqnarray}
&\Psi_k^{(0)}(x,t,\h)=\exp{\Big[-\dac{i}{\h}
\Bigl[\Big(\frac{v^2}{2m}+\tvk V_0\Big)t}+\cr
&+\tvk V_0\dac{\h(2k+1)}{2m\,\Im b}\dil_0^t|C(\tau)|^2d\tau\Bigr)
\Big]\Phi_k^{(0)}(x,t,\h).\label{gauss14}
\end{eqnarray}
The functions (\ref{gauss14}) are approximate, accurate to $O(\h^{3/2})$,
solutions of the Hartree type equation (\ref{gauss1}). However, since
for the linear combination
\begin{equation}
\Phi(x,t)=c_1\Phi_k^{(0)}(x,t)+c_2\Phi_l^{(0)}(x,t)\label{gauss15}
\end{equation}
the condition $\al_\Phi^{(1)}(t,\h)=0$ is not fulfilled, $\Phi(x,t)$ is not
a solution of equation (\ref{gauss7}) and, hence, the linear
superposition principle is invalid for the functions (\ref{gauss14})
even in the class of asymptotical solutions
$\CP_\h^t\big(S(t,\h),P(t),X(t)\big)$ accurate to $O(\h^{3/2})$. Thus,
the presence of the term $\al_{\Phi^{(0)}}^{(1)}(t,\h)$ in equation
(\ref{gauss7}) violates the linear superposition principle
(\ref{gauss15}).

We seek the solution to equation (\ref{gauss1}) in the class
$\CP_\h^t\big(S(t,\h),Z(t,\h)\big)$, i.e., localize the
solution asymptotically in the neighborhood of the trajectory
$z=Z(t,\h)$ depending explicitly on parameter $\h$.  With that, the
estimates (\ref{bbst1.10}) remain valid.  Let us take the dependence of
$Z(t,\h)$ on the parameter $\h\to0$ such that the equation for the
function $\Phi^{(0)}(x,t,\h)$ be linear.  For doing this, we subject
the functions $X(t,\h)$ and $P(t,\h)$ to the equations
\begin{equation}
\left\{\begin{array}{l}
\dot{p}=\dac{2\tvk
V_0}{\gamma^2}\al_{\Phi^{(0)}}^{(1)}(t,\h)+\frac{1}{\gamma^2}
\al_{\Phi^{(0)}}^{(3)}(t,\h),\\[12pt]
\dot{x} =p/m, \end{array}\right. \label{gauss16}
\end{equation}
and the functions $C(t)$ and $B(t)$ to the equations
\begin{equation}
\left\{\begin{array}{l}
\dot B=\dac{\tvk V_0}{\gamma^2}C+\frac{3}{4\gamma^2}
\al_{\Phi^{(0)}}^{(2)}(t,\h)C,\\[12pt]
\dot C=B/m. \end{array}\right. \label{gauss17}
\end{equation}

The function $\Phi^{(0)}(x,t,\h)$ will then satisfy the equation
\begin{equation}
\hat L_0\Phi^{(0)}=0.\label{gauss18}
\end{equation}
Unlike equations (\ref{gauss7})--(\ref{gauss9}), equations
(\ref{gauss16})--(\ref{gauss18}) are dependent.  Note that, within the
accuracy under consideration, the principal term of the asymptotic will
not change if equations (\ref{gauss16}) and (\ref{gauss17})
are solved accurate to $O(\h^{3/2})$ and $O(\h)$, respectively.  Then
equations (\ref{gauss16}) become
\begin{equation}
\left\{\begin{array}{l}
\dot{p}=\dac{2\tvk V_0}{\gamma^2}\al_{\Phi^{(0)}}^{(1)}(t,\h)\\[12pt]
\dot{x} =\dac{p}{m}, \end{array}\right. \label{gauss19}
\end{equation}
and equations (\ref{gauss17}) coincide with
equations (\ref{gauss9}) and their solution has the form
(\ref{gauss10}).  Equation (\ref{gauss18}) is linear and its general
solution can be represented as an expansion over a complete set of
orthonormalized functions $\Phi^{(0)}_k(x,t,\h)$:
\begin{equation}
\Phi^{(0)}(x,t,\h)=\sum_{k=0}^{\iy}c_k\Phi^{(0)}_k(x,t,\h).\label{gauss20}
\end{equation}
Here, $\Phi^{(0)}_k(x,t)$ is determined by expression (\ref{gauss13})
where $X(t)$ and $P(t)$ ought to be replaced by $X(t,\h)$ and
$P(t,\h)$, respectively.  Substitute (\ref{gauss20}) into
(\ref{gauss10}). In view of the properties of Hermite polinomials
\begin{eqnarray*}
\lefteqn{\dil_{-\iy}^{+\iy}\xi H_n(\xi)H_l(\xi)e^{-\xi^2}d\xi =}\\
&=\dil_{-\iy}^{+\iy}\xi\Big[\frac12 H_{n+1}(\xi)+n
H_{n-1}(\xi)\Big]H_l(\xi)e^{-\xi^2}d\xi=\frac12\de_{n+1,l}+n\de_{n-1,l},
\end{eqnarray*}
we obtain
\begin{eqnarray*}
\lefteqn{\al_{\Phi^{(0)}}^{(1)}(t,\h)=
\frac{\sqrt{\h}|C(t)|}{m\,\Im b}\sum_{n=0}^{\iy}\Big(
\frac12\de_{n+1,l}+n\de_{n-1,l}\Big)c_nc_l^{*}=}\\
&&=\dac{\sqrt\h |C(t)|}{m\,\Im b}\sum_{n=0}^{\iy}
\Big(\frac12c_{n+1}^*+nc_{n-1}^*\Big)c_n.
\end{eqnarray*}
Equations (\ref{gauss19}) will then take the form
\[ \left\{\begin{array}{l}
\dot{p}=\Theta_1\sqrt{\h}\cdot|C(t)|\\[12pt]
\dot{x} =\dac{p}{m}, \end{array}\right. \]
where
\begin{equation}
\Theta_1=\frac{\tvk V_0}{m\gamma^2\Im b}
\sum_{n=0}^{\iy}\Big(\frac12c_{n+1}^{*}+nc_{n-1}^{*}\Big)c_n .
\label{gauss21}
\end{equation}
Integration of the obtained equations yields
\[ \left\{\begin{array}{l}
P(t,\h)=m\dot X(t,\h),\\
X(t,\h)=\sqrt{\h}\dac{\Theta_1}{m}\dil_0^t d\tau\dil_0^\tau
|C(s)|ds+\frac{p_0}{m}t+x_0.\end{array}\right. \]
As a result, the principal term of the asymptotic can be represented in
the form
\begin{eqnarray}
\lefteqn{\Psi^{(0)}(x,t,\h) =}\cr
&\displaystyle=\exp\Big\{-\frac{i}{\h}\Bigl[\dil_0^t
\Big(P(\tau,\h)\dot X(\tau,\h)-\frac{P^2(\tau,\h)}{2m}+\h\Theta_2
|C(\tau)|^2\Big)d\tau\Bigr]\Big\}\Phi^{(0)}(x,t,\h), \label{gauss22}
\end{eqnarray}
where
\begin{equation}
\Theta_2=\frac{\tvk V_0}{2m\Im b}\sum_{n=0}^{\iy}\Big[\frac14
c_{n+2}^*+(n+\frac12)c_{n}^{*}+(n^2-n)c_{n-2}^{*}\Big]c_n.\label{gauss23}
\end{equation}
It follows that function (\ref{gauss22}) depends on $\Theta_1$ and
$\Theta_2$ as on parameters:
\[ \Psi^{(0)}(x,t,\h) =\Psi^{(0)}(x,t,\h,\Theta_1,\Theta_2), \]
Here, $\Theta_1$ and $\Theta_2$ are determined by the sets of equations
(\ref{gauss21}) and (\ref{gauss23}), respectively.

Let us consider the Cauchy problem for equation (\ref{gauss1}):
\begin{equation}
\left\{\begin{array}{l}
\Psi_1\Bigr|_{t=0}=\Psi_{10}(x);\quad\Psi_2\Bigr|_{t=0}=\Psi_{20}(x),\\ [6pt]
\Psi_3\Bigr|_{t=0}=\Psi_{30}(x)=G_1\Psi_{10}(x)+G_2\Psi_{20}(x),\\  [6pt]
\Psi_{k}(x)\in\CP_\h^t, \end{array}\right.\label{gauss24}
\end{equation}
where $G_k=\const$.  Denote by $\Psi_{k}(x,t,\h,\Theta_1^k,\Theta_2^k)$
the principal term of the asymptotic solution to equation (\ref{gauss1}),
satisfying the initial conditions (\ref{gauss24}).  Then from the explicit
form of function (\ref{gauss14}) follows
\begin{equation}
\Psi_{3}(x,t,\h,\Theta_1^3,\Theta_2^3)=G_1\Psi_{2}(x,t,\h,\Theta_1^3,
\Theta_2^3)+G_2\Psi_{2}(x,t,\h,\Theta_1^3,\Theta_2^3).\label{gauss25}
\end{equation}
Relationship (\ref{gauss25}) represents the nonlinear superposition
principle for the asymptotical solutions of equation (\ref{gauss1}) in
the class $\CP_\h^t\big(S(t,\h),Z(t,\h)\big)$.

\vspace{2cm}
The work was supported in part by the Russian Foundation for
Basic Research (Grant No. 00-01-00087).
\vspace{2cm}

\section*{Appendix A.  The set of equations in variations}
\addcontentsline{toc}{section}{Appendix A.
The set of equations in variations}

\def\thesection{\rm A}
\setcounter{subsection}{0}
\setcounter{equation}{0}
\setcounter{teo}{0}
\setcounter{demo}{0}
\setcounter{rem}{0}

We already mentioned that to construct solutions to equation (\ref{kak5})
in the class $\CP^t_\h$, it is necessary to find solutions to the
equations in variations (\ref{kak9}) and to the type Riccati matrix equation
(\ref{kak17}).  Let us show that the solutions of the Riccati type
matrix equation can be completely expressed through the solutions of
the equations in variations $a(t)$.

Let us present the $2n$-space vector $a(t)$ in the form
\[ a(t,\Psi_0)=(\vec W(t,\Psi_0),\vec Z(t,\Psi_0)), \]
where the $n$-space vector $\vec W(t)=\vec W(t,\Psi_0)$ is the ``momentum''
part and $\vec Z(t)=\vec Z(t,\Psi_0)$ is the ``coordinate'' part of the
solution of the equations in variations.  Thus we can write the
latter as
\begin{equation}
\left\{\begin{array}{l}
\dot{\vec W}=-\FH_{xp}(t,\Psi_0)\vec W-\FH_{xx}(t,\Psi_0)\vec Z,\\
\dot{\vec Z}=\FH_{pp}(t,\Psi_0)\vec W+\FH_{px}(t,\Psi_0)\vec Z.
\end{array}\right.\label{kak18}
\end{equation}
The set of equations (\ref{kak18}) is called a {\em set of equations in
variations in vector form}. Denote by $B(t)$ and $C(t)$ the $n\times n$
matrices composed of the ``momentum'' and ``coordinate'' parts of the
solution of the equations in variations:
\[ B(t)=(\vec W_1(t),\vec W_2(t),\dots,\vec W_n(t)), \qquad
C(t)=(\vec Z_1(t),\vec Z_2(t),\dots,\vec Z_n(t)). \]
The matrices $B(t)$ and $C(t)$ satisfy the set of equations
\begin{equation}
\left\{\begin{array}{l} \dot B=-\FH_{xp}(t,\Psi_0)B-\FH_{xx}(t,\Psi_0)C,\\
\dot C=\FH_{pp}(t,\Psi_0)B+\FH_{px}(t)C,\end{array}
\right.\label{kak19}
\end{equation}
which is called a set of equations in variations (\ref{kak9}) in matrix form.

Let us consider some properties of the solutions of this set of
equations, which determine the explicit form of the asymptotical
solution of the Hartree type equation and its approximate evolution
operator.

\begin{rem} The set of equations in variations (\ref{kak9}) is a set of
linear Hamiltonian equations with the Hamiltonian function
\[ H(a,t)=\frac 12\lan a,\FH_{zz}(t)a\ran, \qquad a\in{\Bbb C}^{2n}. \]
\end{rem}

The complex number $\{a_1,a_2\}=\lan a_1,Ja_2\ran$ is called a skew-scalar
product of the vectors $a_1$ and $a_2$, $a_k\in{\Bbb C}^{2n}$.

Obviously, the skew-scalar product is antisymmetric:
\[ \{a_1,a_2\}=-\{a_2,a_1\}. \]

\begin{demo} \label{dem1}\rm The skew-scalar product $\{a_1(t),a_2(t)\}$ of
the two solutions, $a_1(t)$ and $a_2(t)$, of equations in
variations (\ref{kak9}) is invariable in time, i.e.,
\begin{eqnarray}
& \{ a_1(t),a_2(t)\}=\{ a_1(0),a_2(0)\}=\const,\\ \label{kak10}
& \{ a_1(t),a_2^*(t)\}=\{ a_1(0),a_2^*(0)\}=\const.\label{kak10a}
\end{eqnarray}
\end{demo}
This statement can be checked immediately by differentiating the
skew-scalar product $\{a_1(t),a_2(t)\}$ with respect to $t$:
\begin{eqnarray*}
& \dac d{dt}\{a_1(t),a_2(t)\}=\lan\dot a_1(t),Ja_2(t)\ran+
\lan a_1(t),J\dot a_2(t)\ran=\\
& =\lan J\FH_{zz}(t)a_1(t),Ja_2(t)\ran+\lan a_1(t),JJ\FH_{zz}(t)
a_2(t)\ran=\\
& =\lan a_1(t),\FH_{zz}(t)a_2(t)\ran-\lan a_2(t),\FH_{zz}(t)a_2(t)\ran
=0.
\end{eqnarray*}
Here, we have made use of the fact that $J^2=-{\Bbb I}_{2n\times 2n}$
and $J^t=-J$.  Relationship (\ref{kak10a}) follows from (\ref{kak10})
since $a_2^*(t)$ is also a solution of the equations in
variations in view of the fact that these equations are real and linear.

For the set of equations in variations in matrix form, Statement
\ref{dem1}\ becomes

\begin{demo} \label{dem2} \rm The matrices
\begin{eqnarray}
& D_0=\dac 1{2i}[C^+(t)B(t)-B^+(t)C(t)],\label{B.6}\\
& \tilde D_0=C^t(t)B(t)-B^t(t)C(t),\label{B.7}
\end{eqnarray}
where the matrices $B(t)$ and $C(t)$ are arbitrary solutions of the set
of equations in variations (\ref{kak19}), are invariable in time, and
so we have
\[ D_0=[C^+(0)B(0)-B^+(0)C(0)]/(2i), \quad
\tilde D_0=C^t(0)B(0)-B^t(0)C(0). \]
\end{demo}

The relation of the matrices $B(t)$ and $C(t)$ to the matrix $Q(t)$
and, in view of (\ref{kak14}), to the function $\phi_1(t)$ yields

\begin{demo} \label{dem3}\rm Let the $n\times n$-matrices $B(t)$ and
$C(t)$ be solutions to equations in variations (\ref{kak9}). Then,
if $\det\,C(t)\ne 0$, $t\in[0,T]$, the matrix $Q(t)=B(t)C^{-1}(t)$
satisfies the Riccati matrix equation (\ref{kak17}).
\end{demo}
Actually, in view of
\[ \dot C^{-1}(t)=-C^{-1}(t)\dot C(t) C^{-1}(t) \]
and since from $C^{-1}(t)C(t) ={\Bbb I}$ follows
$\dot C^{-1}(t)C(t)+C^{-1}(t)\dot C(t)=0$, we have
\begin{eqnarray*}
\dot Q&=& \dot B(t) C^{-1}(t)+ B(t) \dot C^{-1}(t)=\dot B(t) C^{-1}(t)-
Q(t)\dot C (t)C^{-1}(t)=\\
&=&[-\FH_{xp}(t)B -\FH_{xx}(t)C]C^{-1}
- Q[\FH_{pp}(t)B+\FH_{px}(t)C]C^{-1}=\\
&=&-\FH_{xp}(t)Q-\FH_{xx}(t)-Q\FH_{pp}(t)Q-Q\FH_{px}(t),
\end{eqnarray*}
Q.E.D.

A similar property is also valid for the matrix $Q^{-1}(t)$:
\begin{equation}
-\dot Q^{-1} +Q^{-1}\FH_{xx}(t)Q^{-1}+\FH_{px}(t)Q^{-1}+
Q^{-1}\FH_{xp}(t)+\FH_{pp}(t)=0.
\end{equation}

\begin{demo} \label{dem4}\rm If at the time zero the matrix $Q(t)$ is
symmetrical [$Q(0)=Q^t(0)$ at $t=0$], it is symmetrical at any time
$t\in[0,T]$ [i.e., $Q(t)=Q^t(t)$]. Here, $A^t$ denotes the transpose to
the matrix $A$.
\end{demo}
Actually, from (\ref{kak17}) follows
\[\dot Q^t +\FH^t_{xx}(t)+\FH^t_{px}(t)Q^t+Q^t\FH^t_{xp}(t)
+Q^t\FH^t_{pp}(t)Q^t=0, \]
since we have
\[ \FH_{xx}=\FH^t_{xx},\quad \FH_{pp}=\FH^t_{pp},\quad\FH_{px}=\FH^t_{xp}. \]
Hence, the matrix $Q^t(t)$ satisfies equation (\ref{kak17}) with the
same initial conditions as the matrix $Q(t)$ since, as agreed, the
matrix $Q(0)$ is symmetrical.  The validity of the statement follows
from the uniqueness of the solution to the Cauchy problem.

\begin{demo} \label{dem5}\rm The imaginary parts of the matrices $Q(t)$ and
$Q^{-1}(t)$ can be presented in the form
\begin{eqnarray}
&&\Im\,Q(t)=(C^+(t))^{-1}D_0C^{-1}(t),\label{ros24}\\
&&\Im\,Q^{-1}(t)=-(B^{-1}(t))^{+}D_0B^{-1}(t).\label{ros25}
\end{eqnarray}
Here, the matrix $D_0$ is defined by relationship (\ref{B.6}).
\end{demo}
Actually, by definition we have
\begin{eqnarray*}
\Im\,Q(t)&=&\dac i2[Q^+(t)-Q(t)]=
\dac i2\{[B(t)C^{-1}(t)]^+-B(t)C^{-1}(t)\} ={}\\
&=&\dac i2 [C^+(t)]^{-1}[B^+(t)C(t)-C^+(t)B(t)] C^{-1}(t)=
[C^+(t)]^{-1}D_0C^{-1}(t),
\end{eqnarray*}
Q.E.D.  Similarly, relationship (\ref{ros25}) can be proved.

\begin{demo} \label{dem6}\rm Let the matrix $Q(t)$ be definite and
symmetrical and the components of the vector $\vec y\,{}_j^t(t)$,
$j=\overline{1,n}$ be the row elements of the matrix $C^{-1}(t)$
(\ref{B.14}). Then the vectors $\vec y\,{}_j^t(t)$ satisfy the set of
equations
\begin{equation}
\dot{\vec y}=-\CH_{xp}(t) \vec y -Q(t)\CH_{pp}(t)\vec y.\label{B.20}
\end{equation}
\end{demo}
Actually, we have
\[ \dot C^{-1}=-C^{-1}\dot C C^{-1}.\]
and hence
\begin{equation}
\dot C^{-1}= -C^{-1}[ \CH_{pp}(t) Q(t) +\CH_{px}(t)].\label{B.21}
\end{equation}
Transposing relationship (\ref{B.21}) for the vectors $\vec y(t)$
(\ref{B.14}), we obtain equation (\ref{B.20}), Q.E.D.

\begin{rem} If the matrix
\[ Q^{-1}(t) = C(t) B^{-1}(t) \]
is definite and symmetrical, then the matrix $B^{-1}$ satisfies the equation
\begin{equation}
\dot B^{-1}=B^{-1}[ \CH_{xx}(t) Q^{-1}(t) +\CH_{xp}(t)].\label{B.22}
\end{equation}
{\rm The proof is similar to that of Statement \ref{dem6}.}
\end{rem}

\begin{demo}\label{dem7}\rm If the matrix $D_0$ (\ref{B.6}) is positive
definite, the relation
\begin{equation}
2i B^{-1}(t)\CH_{xx}(t)(B^{-1}(t))^t = \frac d {dt} [D_0^{-1} B^+(t)
(B^t(t))^{-1}]\label{B.24}
\end{equation}
is valid.
\end{demo}
Actually, from (\ref{ros25}) follows
\begin{eqnarray*}
\lefteqn{B^{-1}(t)\CH_{xx}(t)(B^{-1}(t))^t=}\\
&&=\displaystyle\frac i 2 D_0^{-1}B^+(t)\big[Q^{-1}(t)-(Q^*(t))^{-1}\big] \CH_{xx}(t)(B^{-1}(t))^t=\\
&&=\frac i 2 D_0^{-1}\big[B^{-1}(t) \big(\CH_{xx}(t)C(t)B^{-1}(t)-\CH_{xx}(t)
C^+(t)(B^{-1}(t))^*+\\
&&
+\CH_{px}(t) B(t)B^{-1}(t) - \CH_{px}(t)B^*(t)(B^{-1}(t))^*
\big)(2i\Im~Q^{-1}(t))B^*(t)D_0^{-1}\big]^t=\\
&&=\frac i 2 D_0^{-1}\big[B^{-1}(t)\dot B(t)B^{-1}(t) B^*(t)-B^{-1}(t)
\dot B^*(t)\big]^t=\\
&&=\frac i 2 D_0^{-1}\Big[\frac d{dt}B^{-1}(t) B^*(t)\Big]^t,
\end{eqnarray*}
Q.E.D.

\begin{demo}\label{dem8}\rm If for equations in variations (\ref{kak19})
the Cauchy problem is formulated as
\begin{equation}
\tilde B(t) |_{t=s} =B_0, \quad \tilde C(t)|_{t=s} =0, \quad
B_0^t=B_0,\label{B.27}
\end{equation}
then the relation
\begin{equation}
\int\limits_s^t\tilde B^{-1}(\tau) \CH_{xx} (t) (\tilde B^{-1}
(\tau))^t d\tau=(B_0^{-1})^t \la_2(\De t,\Psi_0) \la_4^{-1}(\De t,\Psi_0)
(B_0^{-1})^t,\label{B.28}
\end{equation}
is valid. Here, $\la_k(\De t,\Psi_0)$, $k=\overline{1, 4}$ denote the
$n\times n$ matrices being blocks of the matriciant of the set of equations
in variations (\ref{kak9})

\begin{equation}
A(t,\Psi_0)=\left(\begin{array}{l}
\la_4^t(t,\Psi_0)~~\la_2^t(t,\Psi_0)\\
\la_3^t(t,\Psi_0)~~\la_1^t(t,\Psi_0)\end{array} \right),
\qquad A(0,\Psi_0)={\Bbb I}_{2n\times 2n}.\label{A.13}
\end{equation}
\end{demo}
Let us consider an auxiliary Cauchy problem formulated as
\begin{equation}
B(t,\epsilon)|_{t=s}=B_0,\quad C(t,\epsilon)|_{t=s}=\epsilon{\Bbb I},
\quad {\Bbb I}=\|\de_{k,j}\|_{n\times n}.\label{B.29}
\end{equation}
Obviously, we have
\[ \lim_{\epsilon \to 0} B(t,\epsilon)=\tilde B(t), \quad
\lim_{\epsilon \to 0} C(t,\epsilon)=\tilde C(t) \]
and
\[ D_0(\epsilon) =\frac \epsilon {2i} (B_0 -B_0^*).\]
We assume that the matrix $D_0(\epsilon)$ is symmetrical and positive
definite for $\epsilon \ne 0$. Hence, we may use relationship
(\ref{B.24}) and then obtain
\begin{eqnarray}
\lefteqn{\dil_s^t B^{-1}(\tau,\epsilon) \CH_{xx}(\tau)\big(B^{-1}
(\tau,\epsilon)\big)^td\tau =}\cr
&&-\dac 1 \epsilon (B_0 -B_0^*)^{-1}
B^+(\tau,\epsilon) ( B^{-1} (\tau,\epsilon))^t|_s^t.\label{B.30}
\end{eqnarray}
In view of (\ref{A.13}), we have
\[ B(t,\epsilon) = \la^t_4(\De t) B_0 -\epsilon\la^t_2(\De t), \]
and, hence,
\[ B^{-1} (t,\epsilon)=\big(1+\epsilon B^{-1}_0(\la_4^{-1}(\De t))^t\la^t_2
(\De t)\big)B^{-1}_0(\la_4^{-1}(\De t))^t +O(\epsilon^2).\vspace{-1.5ex} \]
Then we obtain
\begin{eqnarray*}
\lefteqn{\lim_{\epsilon\to0}\frac 1 \epsilon B^+(\tau,\epsilon)(B^{-1}
(t,\epsilon))^t|_s^t=\big[\lim_{\epsilon\to 0}[B^{-1}(t,\epsilon)
B^*(t,\epsilon)-B_0^{-1}B_0^*]\big]^t=}\\
&&\hspace{-18pt}=\lim_{\epsilon\to 0}\frac 1 \epsilon\{B_0^{-1}B_0^*\!-\!
\epsilon B_0^{-1}(\la_4^{-1}(\De t))^t\la^t_2(\De t)(1\!-\!B_0^{-1}B_0^*)\!
-\!B_0^{-1}B_0^*\!+\!O(\epsilon^2)]\}^t\!=\\
&&=-(B_0-B_0^*)(B_0^{-1})^t\la_2(\De t)\la_4^{-1}(\De t)(B_0^{-1})^t.
\end{eqnarray*}
Substitution of the obtained expression into (\ref{B.30}) yields (\ref{B.28}).

\begin{demo}\label{dem9}\rm If the matrix $D_0$ (\ref{B.6}) is positive
definite and symmetrical and the matrix $\tilde D_0$ (\ref{B.7}) is zero,
the following relationships are valid:
\begin{eqnarray}
\lefteqn{C^*(t) D_0^{-1}B^t(t)-C(t)D_0^{-1}B^+(t)=}\cr
& =B(t)D_0^{-1}C^+(t) - B^*(t)D_0^{-1}C^t(t)=
2i{\Bbb I}_{n\times n}, \label{B.15}\\
\lefteqn{C^*(t) D_0^{-1}C^t(t)-C(t)D_0^{-1}C^+(t)=}\cr
& =B(t)D_0^{-1}B^+(t) - B^*(t)D_0^{-1}B^t(t)= 0_{n\times n}.\label{B.16}
\end{eqnarray}
\end{demo}
Let us consider an auxiliary matrix $T(t)$ of dimension $2n\times2n$
\begin{equation}
T(t)=\frac 1{\sqrt 2} \left( \begin{array}{lr} D_0^{-1/2} C^t(t) &
-D_0^{-1/2} B^t(t) \\ D_0^{-1/2} C^+(t)& -D_0^{-1/2} B^+(t)
\end{array}\right)\label{B.17}
\end{equation}
and find its inverse matrix.  Direct checking makes us convinced that
\begin{equation}
T^{-1}(t)=-\frac i{\sqrt 2} \left( \begin{array}{lr}
-(B(t)D_0^{-1/2})^* & B(t)D_0^{-1/2} \\
-( C(t)D_0^{-1/2})^*& C(t)D_0^{-1/2}\end{array}\right).\label{B.18}
\end{equation}
Actually, we have
\begin{eqnarray*}
\lefteqn{T(t)T^{-1}(t)=}\\
&\hspace{-12pt}-\displaystyle\frac i 2\!\!\left(\!\!
\begin{array}{lr}
  -[C^t(t)B^*(t)-B^t(t)C^*(t)](D_0^{-1})^*\!\! &\!\!
  [C^t(t)B(t)-B^t(t)C(t)]D_0^{-1}\\
- [C^+(t)B^*(t)-B^+(t)C^*(t)](D_0^{-1})^*\!\!&\!\!
  [C^+(t)B(t)-B^+(t)C(t)]D_0^{-1}\end{array}\!\!\!\right)\!=\\
&=-\displaystyle\frac i 2\left(
\begin{array}{lr}-D_0^{-1/2}(2iD_0)^*(D_0^{-1/2})^*&
 D_0^{-1/2}\tilde D_0 D_0^{-1/2}\\
-D_0^{-1/2}\tilde D_0^* (D_0^{-1/2})^* &
 iD_0^{-1/2} D_0 D_0^{-1/2}\end{array} \right)={\Bbb I}_{2n\times 2n}.
\end{eqnarray*}
From the uniqueness of the inverse matrix follows
\[ T(t)T^{-1}(t)=T^{-1}(t)T(t)={\Bbb I}_{2n\times 2n}, \]
i.e.,
\begin{eqnarray}
\lefteqn{T(t)T^{-1}(t)=}\cr
&-\displaystyle\frac i 2 \left(\begin{array}{lr}
-(BD_0^{-1})^*C^t + BD_0^{-1}C^+&(BD_0^{-1})^*B^t + BD_0^{-1}B^+\\
-(CD_0^{-1})^*C^t + CD_0^{-1}C^+&(CD_0^{-1})^*B^t + CD_0^{-1}B^+\end{array}\right)=\cr
&={\Bbb I}_{2n\times 2n}. \label{B.19}
\end{eqnarray}
However, as assigned, we have $D_0^t = D_0$, and from Definition (\ref{B.6})
follows
\[ D_0^+ = -\frac 1{2i}(C^+B-B^+C)^+ =-\frac 1{2i} (B^+C-C^+B) =D_0.\]
We then have $D_0^* =D_0$ and, hence, $(D_0^{-1})^* =D_0^{-1}$.  Then
from (\ref{B.19}) we obtain (\ref{B.15}) and (\ref{B.16}).

The following properties of the solutions to the set of equations in
variations are dramatically important for the construction of quasi-classical
asymptotics in the class of functions $\CP_\h^t(Z(t,\h),S(t,\h))$.

\begin{lem} \label{lem1} Let the matrix $D_0$ be diagonal and positive
definite and $\det C(t)\ne0$. The matrix $\Im Q(t)$ will then be
positive definite as well.
\end{lem}
{\bf Proof.}  Let we have $D_0=\diag\,(\alpha_1,\dots,\alpha_n)$,
$\alpha_j>0$, $j=\overline{1,n}$. Denote by $\vec y_j^t$ the rows of the
matrix $C^{-1}(t)$:
\begin{equation}
C^{-1}(t) =\left( \begin{array}{l} \vec y{}_1^t(t)\\
\vec y{}_2^t(t)\\ \dotfill\\ \vec y{}_n^t(t)
\end{array}\right).\label{B.14}
\end{equation}
Then for an arbitrary complex vector $|\vec p|\ne 0$ we obtain
\begin{equation}
(\vec p)^+ \Im Q(t)\vec p=\sum_{j=1}^n\lan\vec p,\vec y_j(t)\ran^+
\alpha_j\lan\vec y_j(t),\vec p\ran=\sum_{j=1}^n\alpha_j|
\lan\vec p,\vec y_j(t)\ran |^2>0.\label{B.0}
\end{equation}
The above inequality is true since $|\vec y_j(t)|\ne 0$, and
$\alpha_j>0$, $j=\overline {1,n}$. From this inequality, in
view of the arbitrariness of the vector $\vec p\in\BC^n$, $|\vec p|\ne 0$,
follows the validity of the lemma statement.

\begin{lem} The matrix $C(t)$ is nondegenerate $[\det C(t)\ne0]$ if
the matrix $D_0=(2i)^{-1}(C^+(0)B(0)-B^+(0)C(0))$ is positive definite.
\end{lem}
{\bf Proof.}  Let, for some $t_1$, $\det C(t_1)=0$ be valid. Then a vector
$\vec k$, $|\vec k|\ne 0$ exists, such that
\[ C(t_1)\cdot \vec k=0, \quad (\vec k^+ C^+(t_1) =0).\]
Since relationship (\ref{B.7}) is valid for any $t$, then
\[ \vec k^+ D_0 \vec k=\vec k^+\Big\{\frac i2[B^+(t_1)C(t_1)-
C^+(t_1)B(t_1)]\Big\}\vec k=0. \]

As agreed, the matrix $D_0$ is positive definite, and, hence, the above
equality holds only for $|\vec k|=0$. The obtained contradiction proves the
lemma.

\begin{lem} [Liouville's lemma]\label{liuvil} If the matrix $Q(t)$
is continuous, the relation
\begin{equation}
\exp \Big\{-\frac 12 \inl_0^t \Sp~[\FH_{pp}(t)Q(t) +\FH_{px}(t)]
\,dt\Big\}=\sqrt{\frac{\det C(0)}{\det C(t)}}.\label{ros26}
\end{equation}
is valid.
\end{lem}
{\bf Proof.}  From (\ref{kak19}) follows
\begin{equation}
\dot C = [\FH_{pp}(t)Q(t)+\FH_{px}(t)]C,\label{ros27}
\end{equation}
where the matrix $Q(t)$ is a solution of equation (\ref{kak17}) and the
matrices $\FH_{pp}(t)$ and $\FH_{px}(t)$ are continuous.  Then for the
matrix $C(t)$ the Jacobi identity
\[ \det C(t)=[\det C(0)]\exp\inl_0^t\Sp~[\FH_{pp}(t)Q(t)
+\FH_{px}(t)]\,dt \]
is valid.  Raising the left and right parts of the equality to the
power $-1/2$ yields (\ref{ros26}), Q.E.D.

\end{document}